\documentclass[twocolumn]{aastex701}
\usepackage{amsmath}
\usepackage{bigints}
\usepackage{booktabs}
\usepackage{hyperref}
\usepackage{enumitem}
\usepackage{gensymb}

\newcommand\red[1]{\textcolor{red}{#1}}
\usepackage{soul}

\def\eg{{\it e.g.}}

\def\ie{{\it i.e.}}

\defcitealias{franz2025}{F25}
\def\franz25{\citetalias{franz2025}}

\begin{document}

\title{Search For a Counterpart to the Subsolar Mass Gravitational Wave Candidate S251112cm}

\newcommand{\LCO}{\affiliation{Las Cumbres Observatory, 6740 Cortona Drive, Suite 102, Goleta, CA 93117-5575, USA}}

\newcommand{\UCSB}{\affiliation{Department of Physics, University of California, Santa Barbara, CA 93106-9530, USA}}

\newcommand{\KITP}{\affiliation{Kavli Institute for Theoretical Physics, University of California, Santa Barbara, CA 93106-4030, USA}}

\newcommand{\UCD}{\affiliation{Department of Physics, University of California, 1 Shields Avenue, Davis, CA 95616-5270, USA}}

\newcommand{\WIS}{\affiliation{Department of Particle Physics and Astrophysics, Weizmann Institute of Science, 76100 Rehovot, Israel}}

\newcommand{\OKC}{\affiliation{Oskar Klein Centre, Department of Astronomy, Stockholm University, Albanova University Centre, SE-106 91 Stockholm, Sweden}}

\newcommand{\OAPD}{\affiliation{INAF-Osservatorio Astronomico di Padova, Vicolo dell'Osservatorio 5, I-35122 Padova, Italy}}

\newcommand{\Caltech}{\affiliation{Cahill Center for Astronomy and Astrophysics, California Institute of Technology, Mail Code 249-17, Pasadena, CA 91125, USA}}

\newcommand{\GSFC}{\affiliation{Astrophysics Science Division, NASA Goddard Space Flight Center, Mail Code 661, Greenbelt, MD 20771, USA}}

\newcommand{\UMD}{\affiliation{Joint Space-Science Institute, University of Maryland, College Park, MD 20742, USA}}

\newcommand{\UCB}{\affiliation{Department of Astronomy, University of California, Berkeley, CA 94720-3411, USA}}

\newcommand{\TTU}{\affiliation{Department of Physics, Texas Tech University, Box 41051, Lubbock, TX 79409-1051, USA}}

\newcommand{\STScI}{\affiliation{Space Telescope Science Institute, 3700 San Martin Drive, Baltimore, MD 21218, USA}}

\newcommand{\UT}{\affiliation{Department of Astronomy, The University of Texas at Austin, 2515 Speedway, Stop C1400, Austin, TX 78712, USA}}

\newcommand{\IoA}{\affiliation{Institute of Astronomy, University of Cambridge, Madingley Road, Cambridge CB3 0HA, UK}}

\newcommand{\QUB}{\affiliation{Astrophysics Research Centre, School of Mathematics and Physics, Queen's University Belfast, Belfast BT7 1NN, UK}}

\newcommand{\IPAC}{\affiliation{Spitzer Science Center, California Institute of Technology, Pasadena, CA 91125, USA}}

\newcommand{\JPL}{\affiliation{Jet Propulsion Laboratory, California Institute of Technology, 4800 Oak Grove Dr, Pasadena, CA 91109, USA}}

\newcommand{\Southampton}{\affiliation{Department of Physics and Astronomy, University of Southampton, Southampton SO17 1BJ, UK}}

\newcommand{\LANL}{\affiliation{Space and Remote Sensing, MS B244, Los Alamos National Laboratory, Los Alamos, NM 87545, USA}}

\newcommand{\Tsinghua}{\affiliation{Physics Department and Tsinghua Center for Astrophysics, Tsinghua University, Beijing, 100084, People's Republic of China}}

\newcommand{\NAOC}{\affiliation{National Astronomical Observatory of China, Chinese Academy of Sciences, Beijing, 100012, People's Republic of China}}

\newcommand{\Itagaki}{\affiliation{Itagaki Astronomical Observatory, Yamagata 990-2492, Japan}}
\newcommand{\Einstein}{\altaffiliation{Einstein Fellow}}

\newcommand{\Hubble}{\altaffiliation{Hubble Fellow}}

\newcommand{\CfA}{\affiliation{Center for Astrophysics \textbar{} Harvard \& Smithsonian, 60 Garden Street, Cambridge, MA 02138-1516, USA}}

\newcommand{\UA}{\affiliation{Department of Astronomy and Steward Observatory, University of Arizona, 933 North Cherry Avenue, Tucson, AZ 85721-0065, USA}}

\newcommand{\MPA}{\affiliation{Max-Planck-Institut f\"ur Astrophysik, Karl-Schwarzschild-Stra\ss e 1, D-85748 Garching, Germany}}

\newcommand{\DSFP}{\altaffiliation{LSST-DA Data Science Fellow}}

\newcommand{\catalyst}{\altaffiliation{LSST-DA Catalyst Fellow}}

\newcommand{\HCO}{\affiliation{Harvard College Observatory, 60 Garden Street, Cambridge, MA 02138-1516, USA}}

\newcommand{\Carnegie}{\affiliation{Observatories of the Carnegie Institute for Science, 813 Santa Barbara Street, Pasadena, CA 91101-1232, USA}}

\newcommand{\TAU}{\affiliation{School of Physics and Astronomy, Tel Aviv University, Tel Aviv 69978, Israel}}

\newcommand{\Edinburgh}{\affiliation{Institute for Astronomy, University of Edinburgh, Royal Observatory, Blackford Hill EH9 3HJ, UK}}

\newcommand{\Birmingham}{\affiliation{Birmingham Institute for Gravitational Wave Astronomy and School of Physics and Astronomy, University of Birmingham, Birmingham B15 2TT, UK}}

\newcommand{\CIERA}{\affiliation{Center for Interdisciplinary Exploration and Research in Astrophysics, Northwestern University, 1800 Sherman Ave., 8th Floor, Evanston, IL 60201, USA}}
\newcommand{\NUPA}{\affiliation{Department of Physics and Astronomy, Northwestern University, 2145 Sheridan Road, Evanston, IL 60208, USA}}

\newcommand{\Bath}{\affiliation{Department of Physics, University of Bath, Claverton Down, Bath BA2 7AY, UK}}

\newcommand{\CTIO}{\affiliation{Cerro Tololo Inter-American Observatory, National Optical Astronomy Observatory, Casilla 603, La Serena, Chile}}

\newcommand{\Potsdam}{\affiliation{Institut f\"ur Physik und Astronomie, Universit\"at Potsdam, Haus 28, Karl-Liebknecht-Str. 24/25, D-14476 Potsdam-Golm, Germany}}

\newcommand{\INPE}{\affiliation{Instituto Nacional de Pesquisas Espaciais, Avenida dos Astronautas 1758, 12227-010, S\~ao Jos\'e dos Campos -- SP, Brazil}}

\newcommand{\UNC}{\affiliation{Department of Physics and Astronomy, University of North Carolina, 120 East Cameron Avenue, Chapel Hill, NC 27599, USA}}

\newcommand{\Ohio}{\affiliation{Astrophysical Institute, Department of Physics and Astronomy, 251B Clippinger Lab, Ohio University, Athens, OH 45701-2942, USA}}

\newcommand{\AAS}{\affiliation{American Astronomical Society, 1667 K~Street NW, Suite 800, Washington, DC 20006-1681, USA}}

\newcommand{\MMT}{\affiliation{MMT and Steward Observatories, University of Arizona, 933 North Cherry Avenue, Tucson, AZ 85721-0065, USA}}

\newcommand{\Geneva}{\affiliation{ISDC, Department of Astronomy, University of Geneva, Chemin d'\'Ecogia, 16 CH-1290 Versoix, Switzerland}}

\newcommand{\Steward}{\affiliation{Steward Observatory, University of Arizona, 933 North Cherry Avenue, Tucson, AZ 85721, USA}}

\newcommand{\Leiden}{\affiliation{Leiden Observatory, Leiden University, PO Box 9513, 2300 RA Leiden, The Netherlands}}

\newcommand{\PSU}{\affiliation{Department of Astronomy \& Astrophysics, The Pennsylvania State University, University Park, PA 16802, USA}}

\newcommand{\PSUa}{\affiliation{Department of Astronomy \& Astrophysics, The Pennsylvania State University, University Park, PA 16802, USA}}

\newcommand{\PSUb}{\affiliation{Institute for Computational \& Data Sciences, The Pennsylvania State University, University Park, PA 16802, USA}}

\newcommand{\PSUc}{\affiliation{Institute for Gravitation and the Cosmos, The Pennsylvania State University, University Park, PA 16802, USA}}

\newcommand{\IAIFI}{\affiliation{The NSF AI Institute for Artificial Intelligence and Fundamental Interactions, USA}}

\newcommand{\JHU}{\affiliation{Department of Physics and Astronomy, Johns Hopkins University, 3400 North Charles Street, Baltimore, MD 21218, USA}}

\newcommand{\Utah}{\affiliation{Department of Physics \& Astronomy, University of Utah, Salt Lake City, UT 84112, USA}}

\newcommand{\UIUC}{\affiliation{Department of Astronomy, University of Illinois, 1002 W. Green St., Urbana, IL 61801, USA}}

\newcommand{\Maryland}{\affiliation{Department of Astronomy, University of Maryland, College Park, MD 20742-2421, USA}}

\newcommand{\keck}{\affiliation{W.~M.~Keck Observatory, 65-1120 M\=amalahoa Highway, Kamuela, HI 96743-8431, USA}}

\newcommand{\cbpf}{\affiliation{Laboratório de Inteligência Artificial, Centro Brasileiro de Pesquisas Físicas, 138 Rua Dr. Xavier Sigaud 150, CEP 22290-180, 139 Rio de Janeiro, RJ, Brazil}}

\newcommand{\UFRJ}{\affiliation{Instituto de Física, Universidade Federal do Rio de Janeiro (UFRJ), Caixa Postal 68528, 21941-972 Rio de Janeiro, Brazil}}

\newcommand{\Monash}{\affiliation{School of Physics and Astronomy, Monash University, Clayton, Victoria 3800, Australia}}

\newcommand{\UCSD}{\affiliation{Department of Astronomy \& Astrophysics, University of California, San Diego, 9500 Gilman Drive, MC 0424, La Jolla, CA 92093-0424, USA}}
\newcommand{\Northwestern}{\affiliation{Department of Physics and Astronomy, Northwestern University, Evanston, IL 60208, USA}}

\newcommand{\OzGrav}{\affiliation{OzGrav: The ARC Centre of Excellence for Gravitational Wave Discovery, Clayton, Victoria 3800, Australia}}
\author[0000-0001-7815-7604, gname=Nicholas, sname=Vieira]{Nicholas Vieira}
\email[show]{nicholas.vieira@northwestern.edu}
\CIERA

\author[orcid=0000-0003-4537-3575, gname=Noah, sname=Franz]{Noah Franz}
\email{nfranz@arizona.edu}
\UA

\author[0000-0001-8073-8731, gname=Bhagya, sname=Subrayan]{Bhagya Subrayan}
\email{bsubrayan@arizona.edu}
\UA

\author[0000-0002-5740-7747, gname=Charles, sname=Kilpatrick]{Charles~D.~Kilpatrick}
\email{ckilpatrick@northwestern.edu}
\CIERA

\author[0000-0003-4102-380X, gname=David, sname=Sand]{David J. Sand}
\email{dsand@arizona.edu}
\UA

\author[0000-0002-7374-935X, gname=Wen-fai, sname=Fong]{Wen-fai Fong}
\email{wfong@northwestern.edu}
\CIERA
\NUPA

\author[0000-0002-0832-2974, gname=Griffin, sname=Hosseinzadeh]{Griffin Hosseinzadeh}
\email{ghosseinzadeh@ucsd.edu}
\UCSD

\author[0000-0002-8297-2473, gname=Kate, sname=Alexander]{Kate D. Alexander}
\email{kdalexander@arizona.edu}
\UA

\author[0000-0002-4924-444X, gname=Azalee, sname=Bostroem]{K. Azalee Bostroem}
\email{bostroem@arizona.edu}
\UA
\catalyst

\author[0000-0002-9267-6213, gname=Jillian, sname=Rastinejad]{Jillian Rastinejad}
\email{jcrastin@umd.edu}
\altaffiliation{NHFP Einstein Fellow}
\Maryland

\author[0000-0001-8340-3486, gname=Kerry, sname=Paterson]{Kerry Paterson}
\email{paterson@mpia.de}
\affiliation{Max-Planck-Institut f\"ur Astronomie, K\"onigstuhl 17, 69117 Heidelberg, Germany}

\author[0000-0002-4022-1874, gname=Manisha, sname=Shrestha]{Manisha Shrestha}
\email{manisha.shrestha@monash.edu}
\Monash\OzGrav

\author[orcid=0009-0006-0647-636X, gname=Phillip, sname=Noel]{Phillip Noel}
\email{phillipnoel@arizona.edu}
\UA

\author[0000-0001-8833-474X, gname=Phelipe, sname=Darc]{P. Darc}
\email{phelipedarc@gmail.com}
\cbpf\CIERA

\author[orcid=0000-0002-0744-0047, gname=Jeniveve, sname=Pearson]{Jeniveve Pearson}
\email{jenivevepearson@arizona.edu}
\UA

\author[0000-0002-9085-8187, gname=Aysha, sname=Aamer]{Aysha Aamer}
\email{aamer2@arizona.edu}
\UA

\author[0000-0002-1420-3584, gname=Souza, sname=Santos]{A. Souza Santos}
\email{asantos.astro@gmail.com}
\cbpf

\author[0000-0003-3402-6164, gname=Luidhy, sname=Santana-Silva]{Luidhy Santana-Silva}
\email{luidhy@ov.ufrj.br}
\affiliation{Observat\'orio do Valongo, UFRJ, Ladeira do Pedro Ant\^onio, 43 - Centro, Rio de Janeiro - RJ, 20080-090, Brazil}
\cbpf

\author[0000-0003-4383-2969, gname=Clecio, sname=Bom]{Clecio R. Bom}
\email{clecio@debom.com.br}
\cbpf

\author[0000-0003-4553-4033, gname=Regis, sname=Cartier]{Regis Cartier}
\email{rgcartier@gmail.com}
\affiliation{Centro de Astronomía (CITEVA), Universidad de Antofagasta, Avenida Angamos 601, Antofagasta, Chile}

\author[0009-0007-4271-6444, gname=Hemanth, sname=Bommireddy]{Hemanth Bommireddy}\email{hemanth.bommireddy195@gmail.com}
\affiliation{Department of Astronomy, Universidad de Chile, Camino el Observatorio 1515, Las Condes, Santiago, Chile}
\affiliation{Data and Artificial Intelligence Initiative (IDIA), Faculty of Physical and Mathematical Sciences, Universidad de Chile, Santiago, Chile}

\author[0000-0001-8651-8772, gname=\'Osmar, sname=Rodr\'iguez]{\'Osmar Rodr\'iguez}
\email{olrodrig@gmail.com}
\affiliation{Pontificia Universidad Cat\'olica de Chile, Vicu\~{n}a Mackenna 4860, Macul, Santiago, Chile}
\affiliation{Instituto Milenio de Astrof\'isica (MAS), Nuncio Monse\~{n}or S\'otero Sanz 100, Of. 104, Santiago, Chile}

\author[0000-0003-0123-0062, gname=Jennifer, sname=Andrews]{Jennifer E. Andrews}
\email{jennifer.andrews@noirlab.edu}
\affiliation{Gemini Observatory/NSF's NOIRLab, 670 N. A'ohoku Place, Hilo, HI 96720, USA}

\author[0000-0003-4175-4960, gname=Conor, sname=Ransome]{Conor Ransome}
\email{cransome@arizona.edu}
\UA

\author[0000-0002-8099-9023, gname=Vasileios, sname=Paschalidis]{Vasileios Paschalidis}
\email{vpaschal@arizona.edu}
\UA
\affiliation{Department of Physics, University of Arizona, Tucson, AZ 85721, US}

\author[0000-0002-1468-9668, gname=Jay, sname=Strader]{Jay Strader}
\email{straderj@msu.edu}
\affiliation{Center for Data Intensive and Time Domain Astronomy, Department of Physics and Astronomy, Michigan State University, East Lansing, MI 48824, USA}

\author[0000-0002-2215-1841, gname=Aldana, sname=Grichener]{Aldana Grichener}
\email{agrichener@arizona.edu}
\UA

\author[0000-0001-8602-4641, gname=Jonathan, sname=Quirola-V\'asquez]{J. Quirola-V\'asquez}
\email{jonathan.quirolavasquez@ru.nl}
\affiliation{Department of Astrophysics/IMAPP, Radboud University, P.O. Box 9010, 6500 GL, Nijmegen, The Netherlands}

\author[0000-0002-4951-8762, gname=Sergiy, sname=Vasylyev]{Sergiy Vasylyev}
\email{svasylyev@ucsd.edu}
\UCSD

\author[0000-0001-6082-8529, gname=Marcelle, sname=Soares-Santos]{Marcelle Soares-Santos}
\email{marcelle@physik.uzh.ch}
\affiliation{Department of Physics, University of Zurich, Winterthurerstrasse 190, Zurich, 8057, Switzerland}

\author[0000-0003-0528-202X, gname=Collin, sname=Christy]{Collin T. Christy}
\email{collinchristy@arizona.edu}
\UA

\author[0000-0002-9454-1742, gname=Brian, sname=Hsu]{Brian Hsu}
\email{bhsu@arizona.edu}
\UA

\author[0009-0002-1981-1754, gname=D., sname=Fuls]{D. Carson Fuls}\email{fulsdavid@arizona.edu}
\affiliation{Lunar and Planetary Laboratory, University of Arizona, 1629 E. University Blvd., TucsDon, AZ 85721-0092, USA}

\author[0000-0002-7937-6371, gname=Yize, sname=Dong]{Yize Dong}
\email{yize.dong@cfa.harvard.edu}
\CfA

\author[0000-0002-5060-3673, gname=Daniel, sname=Reichart]{Daniel E. Reichart}
\email{reichart@physics.unc.edu}
\affiliation{Department of Physics and Astronomy, University of North Carolina at Chapel Hill, Campus Box 3255, Chapel Hill, NC 27599-3255}

\author[0000-0003-0737-8463, gname=Jonathan, sname=Pineda-Garc\'ia]{Jonathan Pineda-Garc\'ia}
\email{j.pinedagarca@uandresbello.edu}
\affiliation{Universidad Andrés Bello, Facultad de Ciencias Exactas, Departamento de Ciencias Físicas, Instituto de Astrofísica, Av. Fernández Concha 700, Santiago, Chile}

\author[0000-0003-2594-8052, gname=Kathryne, sname=Daniel]{Kathryne J. Daniel}
\email{kjdaniel@arizona.edu}
\UA

\author[0000-0003-0549-3281, gname=Daryl, sname=Janzen]{Daryl Janzen}
\email{daryl.janzen@usask.ca}
\affiliation{Department of Physics and Engineering Physics, University of Saskatchewan, 116 Science Place, Saskatoon, SK S7N 5E2, Canada}

\author[0000-0002-8925-057X, gname=Carl, sname=Fields]{C. E. Fields}
\email{carlnotsagan@arizona.edu}
\UA

\author[0000-0001-6047-8469, gname=Ann, sname=Zabludoff]{Ann Zabludoff}
\email{aiz@arizona.edu}
\UA

\author[0000-0002-7015-3446, gname=Nicolas, sname=Meza]{Nicolas Meza}
\email{nicomezare@gmail.com}
\UCD

\author[0000-0002-5115-6377, gname=Felipe, sname=Olivares~E.]{Felipe Olivares~E.}\email{folivarese@gmail.com}
\affiliation{UKIRT Observatory, Institute for Astronomy, 640 N. A'ohoku Place, University Park, Hilo, Hawai'i 96720, USA}

\author[0000-0002-0956-7949, gname=Kristine, sname=Spekkens]{Kristine Spekkens}
\email{kristine.spekkens@queensu.ca}
\affiliation{Department of Physics, Engineering Physics and Astronomy, Queen's University, 64 Bader Lane, Kingston, ON K7L 2E1, Canada}

\author[0000-0001-6065-7483, gname=Benjamin, sname=Weiner]{Benjamin Weiner}
\email{bjweiner@arizona.edu}
\affiliation{MMT and Steward Observatory, University of Arizona, 933 North Cherry Avenue, Tucson, AZ  85721-0065, USA}

\author[0000-0002-0025-3601, gname=Maia, sname=Williams]{Maia Williams}
\email{maiawilliams2030@u.northwestern.edu}
\CIERA
\NUPA

\author[0000-0002-2575-2618, gname=Alex, sname=Gibbs]{Alex R. Gibbs}\email{gibbs@arizona.edu}
\affiliation{Department of Physics, University of Arizona, Tucson, AZ 85721, US}
\affiliation{Lunar and Planetary Laboratory, University of Arizona, 1629 E. University Blvd., Tucson, AZ 85721-0092, USA}

\author[gname=Frank, sname=Shelly]{Frank Shelly}\email{shellyf@arizona.edu}
\affiliation{Department of Physics, University of Arizona, Tucson, AZ 85721, US}
\affiliation{Lunar and Planetary Laboratory, University of Arizona, 1629 E. University Blvd., Tucson, AZ 85721-0092, USA}

\author[0000-0002-7352-7845, gname=Aravind, sname=Ravi]{Aravind P. Ravi}\email{apazhayathravi@ucdavis.edu}
\UCD

\author[0000-0001-8738-6011, gname=Saurabh, sname=Jha]{Saurabh W. Jha}\email{saurabh@physics.rutgers.edu}
\affiliation{Department of Physics and Astronomy, Rutgers, The State University of New Jersey, Piscataway, NJ 08854, USA
}

\author[0000-0001-8818-0795, gname=Stefano, sname=Valenti]{Stefano Valenti}
\email{stfn.valenti@gmail.com}
\UCD

\author[0000-0002-6703-805X, gname=Joshua, sname=Haislip]{Joshua Haislip}
\email{jhaislip@gmail.com}
\UNC

\author[0000-0003-4580-3790, gname=David, sname=Trilling]{David E. Trilling}
\email{david.trilling@nau.edu}
\affiliation{Department of Astronomy and Planetary Science, Northern Arizona University, PO Box 6010, Flagstaff, AZ 86011 USA}

\newcommand{\todo}{\textcolor{red}{TODO: }\textcolor{red}}
\newcommand{\note}{\textcolor{blue}{NOTE: }\textcolor{blue}}
\setcounter{footnote}{0}

\begin{abstract}

The recent candidate gravitational-wave (GW) alert from a compact object merger involving at least one subsolar mass (SSM) object has prompted questions about their origins. S251112cm is reported by LIGO/Virgo with a false alarm rate of 1 per 6.2 years, nearby luminosity distance $93 \pm 27$ Mpc, and probability of containing a SSM object of 100\%. Such a system, if astrophysical, likely did not involve the supersolar neutron stars or black holes invoked to explain kilonovae. One must then also invoke hitherto unobserved and speculative models to produce SSM mergers which may have electromagnetic (EM) counterparts. We introduce a framework which vets and scores candidate counterparts to SSM GW events to inform follow-up in search of any among the zoo of potential EM transients: kilonovae, kilonovae-within-supernovae, super-kilonovae, or AGN flares from binary black hole mergers. We use a suite of telescopes to perform tiling, galaxy-targeted observations, and photometric/spectroscopic follow-up of promising candidates. In near-real time, we ingest candidates reported by the community, including some of the first observations reported by the Vera C. Rubin Observatory. We vet and score a total of 456 candidates, including 67 from Rubin, but find no likely counterpart. We nonetheless highlight candidates which demonstrate the ability of our framework to distinguish between different transient types and describe strategies to maximize the chances of detecting a counterpart to the next SSM event. Our framework will be implemented in the forthcoming Multimessenger Tool for Rapid Object Vetting and Examination (\texttt{TROVE}).

\end{abstract}



\section{Introduction}

The landmark detection of both gravitational waves (GWs) \citep{ligo_scientific_collaboration_and_virgo_collaboration_gw170817_2017} and electromagnetic (EM) emission (\citealt{abbott_multi-messenger_2017}; see \citealt{margutti_chornock_2021_GW170817_review, nicholl_andreoni_2025_GW170817_review} for reviews) from the binary neutron star (BNS) merger GW170817 ushered in the era of GW + EM multimessenger astronomy. Yet, no new unambiguous GW + EM detections have been made since. This period of quiet has endured despite the LIGO-Virgo-KAGRA (LVK) GW detector network operating during two observing runs (Observing Run 3 (O3), April 2019 - March 2020; Observing Run 4 (O4), April 2023 - November 2025; \citealt{gwtc3, gwtc4_catalog}). Each run has seen substantial increases in detector and search pipeline sensitivity \citep{buikema20_aLIGO_sensitivity_O3, acernese23_Virgo_sensitivity_O3, abac25_GWTC4},  increasing the number of GW events which merit EM follow-up. Multimessenger endeavors have been supported by the advent of tools such as the GW Treasure Map \citep{wyatt20_GWTreasureMap}, \texttt{teglon} \citep{coulter25_GW190425_teglon}, Target and Observation Managers (TOMs; \citealt{street18_TOMs}), and increased adoption of existing tools such as General Coordinates Network (GCN) alerts and the Transient Name Server (TNS). One such tool in active development is the Tool for Rapid Object Vetting and Examination (\texttt{TROVE}; \citealt{franz2025}). \texttt{TROVE} will assist in efficiently allocating finite photometric and spectroscopic resources when searching for counterparts associated with poorly localized events, such as GWs.

These tools, and maximum coordination among the community, will be crucial with the advent of regular operations from the Vera C. Rubin Observatory and its ten-year optical Legacy Survey of Space and Time (LSST) \citep{ivezic19_LSST, andreoni22_VRO_ToO, bianco22_Rubin_LSST}. Moreover, the LVK is planning for a 6-month observing run (IR1) in late 2026\footnote{ \href{https://observing.docs.ligo.org/plan/}{https://observing.docs.ligo.org/plan/}}, and additional observing runs in the coming years \citep{abbott18_LRR}. Tools such as \texttt{TROVE} are timely as we are currently witnessing the possible emergence of a new class of GW events for which our expectations for EM counterparts are not developed: events involving the merger of one or more objects which are subsolar in mass.

The LVK began applying two pipelines (\texttt{GstLAL}; \citealt{tsukada23_GstLAL} and \texttt{MBTA}; \citealt{allene25_MBTA}) to search for GWs from subsolar mass (SSM) compact object mergers in their data stream in real time, and began reporting the probability that a system contains at least one SSM object (\texttt{HasSSM}), in February 2025.\footnote{\href{https://emfollow.docs.ligo.org/userguide/changes.html\#version-26-2025-02-20}{https://emfollow.docs.ligo.org/userguide/changes.html\#version-26-2025-02-20}} These pipelines rely on a new bank of template GW waveforms \citep{hanna2025_SSM-templates} for mergers which include at least one object in the mass range $[0.2, 1.0]~M_{\odot}$, spanning mass ratios $q = m_2 / m_1 = 0.1$~to $1.0$, where $m_2 \leqslant m_1$ (and thus $q \leqslant 1.0$) by convention. Offline searches have been conducted for previous observing runs, through the end of O4a (\citealt{lvk26_SSMsearchO4a}, and references therein), but have yielded no conclusive detections.

The GW event S251112cm was observed on 12 November 2025 at 15:18:45.363443 UTC with contributions to detections from all three operational GW detectors \citep{gcn42650, gcn42690}, with a False Alarm Rate (FAR) of 1 per 6.2 years\footnote{The latter GCN reports a FAR of 1 in 4 years, while the number on GraceDB and in the most recently published localization regions remains at 1 in 6.2 years: \href{https://gracedb.ligo.org/superevents/S251112cm/view/}{https://gracedb.ligo.org/superevents/S251112cm/view/} } and a GW luminosity distance of $93\pm27~\mathrm{Mpc}$. S251112cm was triggered by only the MBTA SSM pipeline as the relevant parameter space is not explored by GstLAL in real time. The signal is reported with a high log-Bayes coherence factor $\ln B = +6.1$ across the three contributing GW detectors, indicating a likely coherent signal. It is also reported with a probability that the system contains at least one SSM object \texttt{HasSSM} = 1.0 and probability that the system contains an object of $1 - 3~M_{\odot}$ (referred to by the LVK as \texttt{HasNS} for SSM events) of just 0.08.  Parameter inference yields a source frame chirp mass $\mathcal{M}$ in the bin $[0.10, 0.87]~M_{\odot}$. These facts make S251112cm the first significant ($\mathrm{FAR} < 0.5~\mathrm{year^{-1}}$, for SSM events) and relatively nearby GW event reported by LVK to involve at least one SSM object. 

Whether SSM GW mergers exist in nature is unknown. But, the LVK reports that S251112cm is a significant GW signal with a high coherence factor across the three GW detectors, warranting further exploration of potential EM counterparts and optimal search strategies. We therefore take the prospect of SSM GW events seriously in this work.

\begin{figure}[!t]
    \centering
    \includegraphics[width=\linewidth]{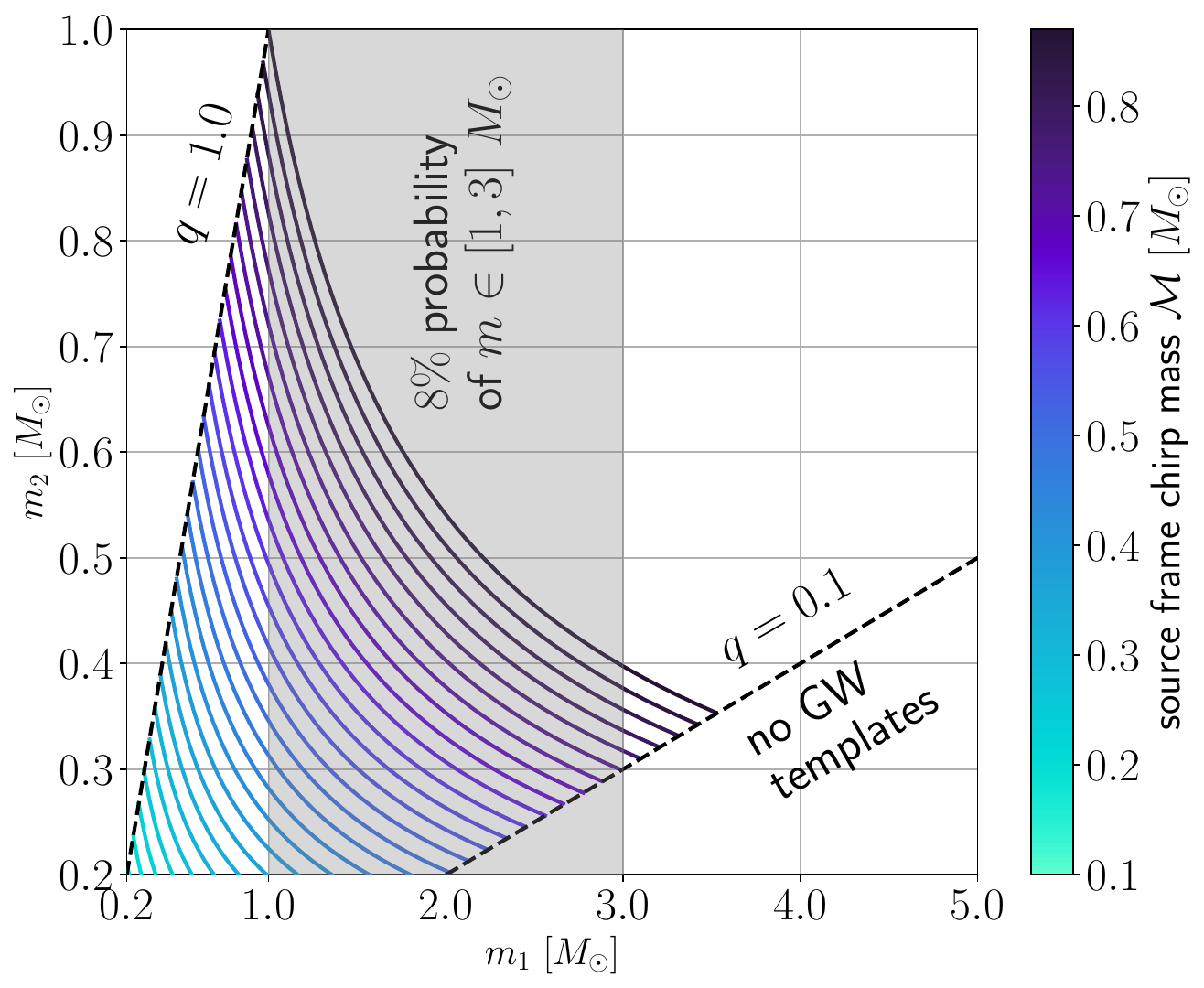}
    \caption{\textbf{{Primary and secondary masses $m_1$ and $m_2$ for varying source frame chirp mass $\mathcal{M}$ in the range $\mathbf{[0.10, 0.87]~M_{\odot}}$}.} The progenitor system $(m_1, m_2)$ for S251112cm may lie along any of the plotted curves. We adopt the LVK convention that $m_1 \geqslant m_2$ such that the mass ratio $q = m_2 / m_1 \leqslant 1.0$. Depending on the chirp mass, $m_1$~may also be sub or supersolar in mass. We truncate for both $m_1~\mathrm{and}~m_2 \leqslant 0.2~M_{\odot}$ and $q \leqslant 0.1$, given the limits of waveform template banks \citep{hanna2025_SSM-templates}. We highlight the range $1 - 3~M_{\odot}$. S251112cm was reported by the LVK with a probability of containing a $1 - 3~M_{\odot}$ object of just 0.08.}
    \label{fig:chirpmassm1m2}
\end{figure}

The progenitor of this novel GW signal, if astrophysical in origin, is unknown. Figure~\ref{fig:chirpmassm1m2}~shows primary and secondary masses $m_1$ and $m_2$ for varying chirp mass in the reported $[0.10, 0.87]~M_{\odot}$ range. As $m_1 \geqslant m_2$ and \texttt{HasSSM} = 1.0, $m_2$ must be subsolar. We truncate this plot for $m_1~\mathrm{or}~m_2 \leqslant 0.2~M_{\odot}$ and mark the region in which $q \leqslant 0.1$ given the limits of the template banks, and highlight in gray the aforementioned range $1 - 3~M_{\odot}$. To explore which $m_1$ are allowed, we write:

\begin{equation}
    m_1 = \mathcal{M} \Big( \frac{1+q}{q^3} \Big)^{1/5} .
\end{equation} 

\noindent For near-symmetric $q \sim 1.0$ and sufficiently small $\mathcal{M}$, $m_1$ may also be subsolar. Considering the limitations of the template banks and low  probability of a $1-3~M_{\odot}$ object being involved, it is most likely that S251112cm was produced by the merger of two SSM objects with such $q \sim 1.0$. But, $m_1 \sim 1.0 - 3.5~M_{\odot}$~and mass ratios approaching $0.1$~are not completely ruled out. 

The merging system could consist of two black holes (BBH), two neutron stars (BNS), or a neutron star-black hole binary (NSBH), but, at least one of the objects must be subsolar in mass. Mergers involving white dwarfs (WDs) or the nondegenerate cores of stars are disfavored as these produce GWs with frequencies of mHz to dHz, outside of LIGO/Virgo bands \citep{maselli_WDWDmergers_2020, grichener25}. Such a SSM neutron star (ssNS) or black hole (ssBH) would be the lightest neutron star or black hole known to date. How such SSM compact objects could be formed remains an open question, as ssNS/ssBHs have not been definitively observed. \citet{doroshenko22_0.77Msun_HESS_J1731-347} propose that the compact object at the center of supernova remnant HESS J1731-347 may have a mass of $0.77^{+0.20}_{-0.17}~M_{\odot}$, but \citet{alford-halpern23} highlight that this measurement is highly distance-dependent and did not analyze all available XMM-Newton data, arguing instead for a supersolar object. The next-lightest known NSs are supersolar: a NICER analysis infers a mass of $1.04^{+0.05}_{-0.03}~M_{\odot}$ for pulsar PSR J1231-1411 (\citealt{salmi24_NICER_PSR_J1231-1411}), and there is evidence for a $1.174 \pm 0.004~M_{\odot}$ NS in the binary pulsar system J0453+1559 (\citealt{martinez2015_lightestNS}, but see \citealt{tauri19_PSRorWD} for an argument that the object is instead a WD). Moreover, it is challenging to explain how standard binary evolution could yield a system of ssNSs/ssBHs. Modern 3D simulations of neutrino-driven core collapse supernovae do not yield remnant NSs lighter than $\sim$$1.17-1.19~M_{\odot}$ \citep{suwa18_SNe_lightNSs, muller25_SNe_lightestNS}. Accretion-induced collapse of WDs similarly yields NSs of $\gtrsim$$1.1-1.2~M_{\odot}$ \citep{nomoto1991_accr-induced-WD-collapse}. We therefore explore alternative formation channels for ssNSs or ssBHs, how they might merge, and the resultant EM counterparts.

Mergers of these objects may produce EM counterparts under specific conditions. In the case of BNS/NSBH mergers in isolation, we expect a kilonova (KN; see \citealt{metzger19_KNe} for a review) powered by the radioactive decay of elements freshly synthesized by rapid neutron capture onto heavy seed nuclei (the $r$-process; \citealt{cowan_rprocess_2021}). Other models propose hitherto unobserved and somewhat speculative transients. \citet{metzger_fragmentation_2024}, \citet{chen_gravitational_2025} and \citet{lerner26_fragmentation} explore scenarios where ssNSs form and merge with each other in the BH accretion disks around collapsars, yielding a ``kilonova-within-a-supernova''. \cite{siegel22_superKN} explore more massive collapsar disks, yielding ``super-kilonovae''. Finally, a BBH merger which occurs in a baryon-rich environment could be EM-bright, with the disks of active galactic nuclei (AGN) being a promising environment (\eg, \citealt{mckernan19_BBHAGN, Kimura2021, Tagawa2024, rodriguez25_AGNflare}). 

Associating a SSM GW event with any EM counterpart would have far-ranging physical implications: on the viability of these GW+EM multimessenger signals, formation channels for SSM compact objects, and the search for primordial black holes and their mergers \citep{ali_haimoud_pBHs_2017, carr_pBHs_2021, crescimbeni24, riajulhaque26}, but also the origins of the heavy elements, the neutron star equation of state \citep{crescimbeni25}, constraining the Hubble constant using combined GW + EM detections, and tests of general relativity. Unsurprisingly, several teams searched for an EM counterpart to S251112cm, yielding a total of 456 candidate counterparts reported in GCN alerts and/or to the TNS in the localization region of S251112cm up to 10 days immediately following the GW signal (see Section~\ref{ssc:scoring-photom} for a justification of this timescale). However, none have been confidently associated with S251112cm. 

Considering this zoo of potential progenitors and EM counterparts, searches for counterparts to SSM events may need to adopt additional strategies beyond those used in searching for counterparts to BNS or NSBH events. We therefore introduce a framework for scoring and ranking candidate EM counterparts to GW events with substantial probability of containing a SSM object. This framework builds on that introduced in \cite{franz2025} (hereafter \franz25) for scoring candidates as kilonova counterparts to BNS/NSBH mergers. We introduce new scores for kilonovae-within-supernovae (KNe-in-SNe), super-kilonovae (super-KNe), and BBH merger-induced AGN flaring. The scores introduced in our framework here will be a part of the forthcoming multimessenger \texttt{TROVE}.


\section{Possible Progenitor Systems Of Subsolar Mass Mergers}\label{sec:possibleprogenitors}

The potential progenitor systems for SSM events fall into two categories: (1) those hosting one or more NSs and (2) BBH systems. Considering the viability of detection for both EM and GW signals, we explore systems which might produce KNe, KNe-in-SNe, and super-KNe in Section~\ref{ssc:EMcounterparts-NS}. We consider BBH mergers which induce AGN flaring in Section~\ref{ssc:EMcounterparts-BBH}. Here, we focus on introducing the potential progenitors; in Section~\ref{sec:scoring}, we quantify the corresponding transient EM signatures and introduce a framework for scoring candidate counterparts based on comparison to these different transients.

\subsection{Neutron Star-Containing Systems}\label{ssc:EMcounterparts-NS}

The S251112cm GW alert is inconsistent with a BNS/NSBH merger involving two supersolar objects, which can generate a KN. \citet{markin23} and \citet{corman26} have explored BNS/NSBH mergers where one component is subsolar and the ensuing KNe, but the existence of such compact objects remains unknown. However, \citet{east_eccentric_2015} propose a scenario in which a highly eccentric NSBH system results in the BH stripping the NS on a first grazing orbit, rendering it subsolar, before later merging. Such a system could, in theory, produce two EM signatures and just one SSM GW signal. More modeling is required to determine the time separation between the EM signatures, so we include these events under the broader umbrella of KNe, which we search for among candidates following \franz25. 

\cite{metzger_fragmentation_2024}, \cite{chen_gravitational_2025} and \cite{lerner26_fragmentation} propose a channel for ssNS formation in the accretion disk of a BH left behind by the collapse of a rapidly-rotating massive star, \ie, a collapsar. In this scenario, gravitational instabilities in the disk lead to fragmentation into clumps which condense to form ssNSs. The requisite conditions may arise if the original star has a mass $\gtrsim$$20~M_{\odot}$ and is rotating with enough angular momentum at collapse to leave behind a $\sim$$1-10~M_{\odot}$~disk around a $\sim$$3-30~M_{\odot}$ BH. The recently-formed ssNSs in the disk may merge with each other, producing a GW inspiral signal. Multiple such mergers can occur if more than two ssNSs are formed. The GWs produced by these mergers may show evidence of substantial eccentricities \citep{wu26_fragmentation_eccentricity}. These mergers ultimately yield some hierarchically-formed object with mass $\lesssim$$2~M_{\odot}$, which finally merges with the central remnant BH, producing one final GW inspiral signal. 

This proposed sequence of collapse and mergers has unique EM signatures. Before any mergers, one expects a long gamma-ray burst (lGRB) which could be detected if beamed towards the observer, and the beginning of a core-collapse supernova (CCSN) \citep{macfaydenwoosley99, woosley06_lGRBSN, dainotti22_GRBSNe}. Among CCSNe, stripped-envelope supernovae (SESNe) and especially SNe Ic-BL are more favored, as they are the only SNe associated with GRBs. The mergers of ssNSs then produce a KN-in-SN \citep{metzger_fragmentation_2024}. These KNe are unlikely to resemble AT2017gfo due to their embedding in the bright SN and the dense accretion disk, but the lightcurves and spectra of the SN may nonetheless be marked by the presence of $r$-process elements \citep{siegel19_collapsar, barnes_2022, patel_effects_2024} or energy injection from a hierarchically-formed millisecond magnetar (\citealt{bucciantini_2012, metzger_piro_2014}).

We also explore the potential for ssNS mergers in disks left behind by more massive (zero-age main sequence mass $\gtrsim$$260~M_{\odot}$; helium core $\gtrsim$$130~M_{\odot}$) stars. \cite{siegel22_superKN} show that for sufficiently high accretion rates, disk winds in such systems unbind as much as $\gtrsim$$50~M_{\odot}$, hosting $\gtrsim$$5-10~M_{\odot}$~of $r$-process material and $\sim$$0.1 - 1.0~M_{\odot}$~of nickel-56 (${}^{56}$Ni). The predicted transient might then resemble a lanthanide-rich KN in color, but evolve on much longer timescales $\gtrsim 1~\mathrm{month}$. \cite{siegel22_superKN} thus term these events super-KNe. If the collapsar disk in this scenario also experiences fragmentation, this fragmentation might yield ssNS mergers and produce GW inspiral signals,. \cite{chen_gravitational_2025} point out that fragmentation might be even easier to achieve for these more massive collapsar disks given higher mass-infall rates. A detailed study of fragmentation in these more massive disks could clarify any differences between expected GW signals. 

We include two important caveats to the KN-in-SN and super-KN schemes. First, we note the absence of other GW signals spatially or temporally associated with S251112cm. If more than two SSM objects were formed in a collapsar disk, one would expect multiple S251112cm-like GW signals within $\sim$hours to $\lesssim 1$ day of each other, though potentially weaker if the objects were less massive. The inspiral and merger of the hierarchically formed product of these merger(s) with the central BH should also be detectable. If such a $\sim$$2~M_{\odot}$ object finally merged with a central remnant BH of $\sim$$20~M_\odot$, this final merger might resemble the $\sim$$23 + 2.6~M_{\odot}$~merger GW190814 \citep{abbott20_GW190814}, detected by LIGO and Virgo at $\sim$$260~\mathrm{Mpc}$ at O3 sensitivities. If a $\sim$$2~M_{\odot}$ object finally merged with the more massive ($\gtrsim$$60~M_{\odot}$) BH expected to be present in super-KNe, such a merger would also lie within existing GW template banks \citep{gwtc4_methods}, though no such $\gtrsim$$60$ + $\sim$$2~M_\odot$ merger has been published by LIGO/Virgo to date \citep{gwtc3, gwtc4_catalog}. Second, neither KNe-in-SNe nor super-KNe have been unambiguously observed. \cite{kasliwal_25ulz_2025} argue that the candidate counterpart SN\,2025ulz to GW event S250818k may have been a KN in a SN\,IIb,\footnote{\cite{kasliwal_25ulz_2025} suggest broadening the definition of super-KNe to refer to any CCSNe hosting some KN-like $r$-process nucleosynthesis, including KNe-in-SNe as explored in \cite{metzger_fragmentation_2024}. We retain the distinction here, for precision.} but see several other arguments against this \citep{franz2025, gillanders_pan-starrs_2025, hall25_AT2025ulz, oconnor25_AT2025ulz, yang26_AT2025ulz, ackley26_AT2025ulz}. We agree that SN\,2025ulz is a normal SN\,IIb. 

Nonetheless, the advent of a potential SSM GW detection requires that we consider a multitude of potential pathways for production of SSM compact objects, their mergers, and predicted EM counterparts. Although the absence of coincident GW signals challenges the KN-in-SN and super-KN models, the S251112cm GW alert is also inconsistent with a canonical supersolar BNS or NSBH merger. We thus search for KNe among candidate counterparts, but also, introduce metrics within \texttt{TROVE} to search for KNe-in-SNe and super-KNe.

\subsection{Binary Black Hole Systems}\label{ssc:EMcounterparts-BBH}

An exciting potential progenitor pathway for an SSM GW event is if one or both compact objects are primordial black hole (PBHs) with masses $< 1M_\odot$. There are few predictions for the EM counterparts to PBH mergers \citep{deng18_PBHs_FRBs, liu20_PBHs_charged}, but we assume that the EM signal hypothesized to originate from massive BBH mergers is mass-scale invariant and may still apply in this low-mass regime. Such BBH mergers may be EM-bright if they occur in an environment rich in baryons and among the most promising environments for this scenario is within the disk of an AGN. In this scenario, a BBH merger yields a kicked remnant which accretes material, producing a hotspot of emission off-center from the AGN before exiting the AGN disk (\citealt{mckernan19_BBHAGN}). Based on predicted temperatures, such a merger could manifest as a UV/optical flare \citep{mckernan19_BBHAGN, Kimura2021, RodrguezRamrez2023,  Tagawa2024, darc25_BBHflares, rodriguez25_AGNflare, mcpike26_McFACTSIV}. Aside from this hotspot, other mechanisms such as the launching of a jet or the interaction of said jet with the AGN disk may also produce transient UV, optical, or X-ray emission; we use the term ``flare'' to broadly refer to any transient EM emission. We can gain some insight into the formation and pre- and post-merger behavior of BBH systems embedded in disks from studies of stellar mass \citep{grishin24_BBH, moncrieff26_BBH, vaccaro26_BBH} and supermassive black hole (SMBH) binary systems \citep{gold14_SMBBH, khan18_SMBBH, bright23_SMBBH, avara24_SMBBH, ennoggi25_SMBBH, manikantan25_SMBBH_eccentricity, manikantan25_SMBBH_MAD}. Systems of ssBH binaries in AGN disks would benefit from similar explorations.

Such a flare may have been seen in AGN J124942.3+344929, tentatively associated with the BBH merger GW190521  (\citealt{graham20_GW190521flare}, but see \citealt{ashton21_GW190521}). However, GW190521 involved a far more massive binary system ($\sim$$85 + 66$~$M_{\odot}$), and the works referenced above also consider more massive systems. This difference matters, as models invoke Bondi-Hoyle-Lyttleton accretion onto the remnant BH and thus generically predict that flare luminosities scale as $L \propto M_{\mathrm{BBH}}^2 = (m_1 + m_2)^2$. A merger with S251112cm-like masses (\eg, $0.5 + 0.5~M_{\odot}$) could therefore be a factor of $\sim$$10^4$ less luminous than the flare potentially associated with GW190521, based on mass alone. 

However, the  luminosity and timescale of the flare also depend on the kick velocity of the remnant, the angle between the kick and the AGN disk midplane, the density of the ambient AGN disk, and the mass of the SMBH at the center of the AGN. Though likely challenging for a BBH merger with S251112cm-like masses to yield a sufficiently luminous flare, there may be select regions of the parameter space allowed by S251112cm, kick configurations, and AGN disks with favorable conditions which combine to produce an observable flare. Associating such a flare with S251112cm would thus place tight constraints on the progenitor and the host AGN itself. We thus also introduce metrics to search for AGN flares as EM counterparts to SSM GW events in \texttt{TROVE}.


\begin{figure*}[t!]
    \centering
    \includegraphics[width=\linewidth]{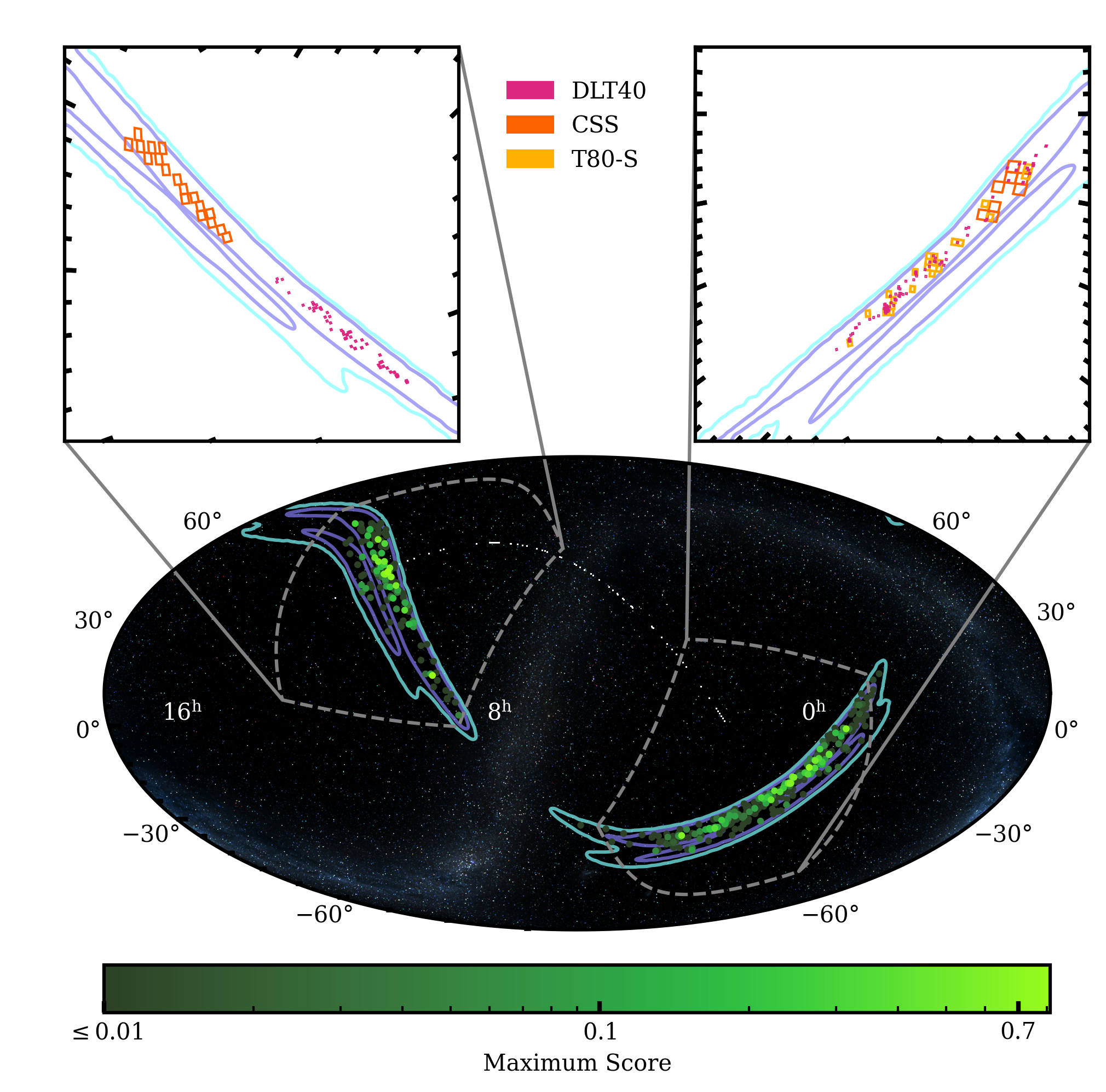}
    \caption{\textbf{GW localization region of S251112cm, with our telescope pointings in search of an optical counterpart, and candidates identified by all teams.} Purple and blue contours denote the 50\% and 90\% confidence localization regions, which span 371 and 1,681 $\mathrm{deg}^2$, respectively. \textit{Top left and right:} Zoomed-in insets show our telescope pointings in search of a counterpart with DLT40 (magenta), CSS (orange), and T80-S (yellow), in the two lobes of the localization region. \textit{Bottom}: Points denote the 456 candidate counterparts reported to the TNS and/or in GCNs within 10 days of the GW signal, color-coded by score, where we select the maximum score among KN, KN-in-SN, and super-KN scores (Section~\ref{sec:scoring}).}
    \label{fig:skymap}
\end{figure*}

\section{Search for an Electromagnetic Counterpart to S251112\lowercase{cm}} \label{sec:obs} 

SSM GW events have unclear origins, progenitors, and EM counterparts. Furthermore, we see only one GW signal in S251112cm, while some of the speculative models discussed here predict multiple GW signals. Despite these important caveats, we begin from the position that S251112cm may be an astrophysical GW signal produced by a merger involving one or more SSM compact objects. We vet and score candidate EM counterparts to S251112cm found during the intensive follow-up campaign for the event.

Several teams searched for an EM counterpart in the localization region of S251112cm in the optical/IR, X-rays, and gamma-rays. Other teams serendipitously contributed to the wealth of candidates by conducting their regularly-scheduled wide-field transient surveys and reporting transients to the TNS. In all, we find 456 candidate counterparts in the 10 days following the event, including 426 in the optical and 30 in X-rays. We impose this cutoff at 10 days, as extending to much later times would yield an infeasible number of serendipitously detected candidates, and the slowest-rising transients we consider (super-KNe) reach at minimum $\sim$50\% of their peak brightness by this time (Section~\ref{ssc:scoring-photom}).

Our own search for a counterpart involved (1) tiling as much of the localization region as possible with larger field-of-view (FOV) instruments, (2) galaxy-targeted observations with smaller-FOV instruments, (3) vetting candidate counterparts with \texttt{TROVE}, and (4) targeted optical and spectroscopic follow-up of promising candidates. The majority of this work was conducted within the Searches After Gravitational waves Using ARizona Observatories (SAGUARO) collaboration \citep{lundquist19_SAGUARO}, which used the SAGUARO TOM \citep{hosseinzadeh24_SAGUARO}\footnote{For other examples of TOMs, see \citet{2018SPIE10707E..11S, vanderWalt2019, Coughlin_2023, 2023PASP..135f4501C}, and \citet{bhtom}.} and a suite of telescopes to search for and characterize candidate EM counterparts to GW events. All photometry used in this work will be made publicly available on the Open mulTiwavelength Transient Event Repository \citep[{\tt OTTER};][]{2026JOSS...11.9516F, 2026ApJ...999..243F}\footnote{\url{https://otter.idies.jhu.edu}}.

\subsection{Search for an Optical Counterpart}\label{ssc:obs-search}

We searched for an optical counterpart to S251112cm using (1) the 1.5 m Mt. Lemmon telescope operated by the Catalina Sky Survey (CSS) in Arizona, (2) the 0.82 m T80-South (T80-S) Telescope at the Cerro Tololo Inter-American Observatory (CTIO) in Chile, and (3) the 0.4 m Panchromatic Robotic Optical Monitoring and Polarimetry Telescopes (PROMPT) \citep{2005NCimC..28..767R} at CTIO and the Meckering Observatory in Australia, as part of the Distance Less Than 40 Mpc (DLT40) Survey. We show all pointings with these telescopes in Figure~\ref{fig:skymap}. All pointings have been uploaded to the GW Treasure Map \citep{wyatt20_GWTreasureMap}.  

During normal operations, the 1.5 m Mt. Lemmon telescope is used by the CSS to discover and characterize solar system objects. It is equipped with a 5 deg$^2$ prime focus imager and observations are taken with no filter, then calibrated to the Gaia $G$-band. When a GW alert of interest is reported, SAGUARO is able to rapidly request observations using the SAGUARO TOM \citep{hosseinzadeh24_SAGUARO} to survey the localization region \citep[see][for a further description of SAGUARO operations]{lundquist19_SAGUARO, paterson21_SAGUARO}. Candidate transients are found by running a difference imaging pipeline \citep{2023zndo...8436113P} based on the algorithm of \cite{zackay16_ZOGY} and a standard machine learning classifier. Our first epoch of observations tiled $\sim$$55$ deg$^2$ of the southern lobe of the S251112cm localization region in this mode, at $\sim 12.8 - 12.9$ hours post-merger.  In addition to triggered observations, the CSS serendipitously observed a further $\sim$$90$ deg$^2$ of the northern lobe during normal solar system search operations at $\sim 43.6- 44.4$ hours post-merger, in which we searched for counterparts as well. In all, SAGUARO searched a total area of $145~\mathrm{deg}^2$, reaching a mean $5\sigma$ limiting magnitude of $\sim$20.7~(AB) mag at 0.5 and 1.8 days post-merger. 

At the same time as our first CSS observations, we used the 1.4$\times$1.4~deg$^{2}$ T80S-Cam on T80-S to observe 23 unique fields in the 90\% localization region, covering approximately 45~deg$^{2}$, in the $i$-band. We reduced all imaging using {\tt photpipe} \citep{Rest05}, following standard reduction procedures as described in \citet{Santos24}. We ran difference imaging on all imaging using Dark Energy Camera (DECam) $i$-band templates to search for transient emission. We achieved an average 3$\sigma$ limiting magnitude for sources in the difference images of $i > 21.2$~mag at 0.5 days post-merger.

Finally, we performed galaxy-targeted observations with DLT40 \citep{Tartaglia18}. DLT40 is designed as a high cadence SN search, with 0.4 m telescopes in Chile, Australia, and Canada\footnote{The Canadian telescope was offline during S251112cm and thus did not play a role in our search.}, each equipped with identical 10$\times$10 arcmin$^2$ imagers. When a GW event of interest is announced, DLT40 performs a prompt galaxy-targeted search with real-time difference imaging and candidate identification, down to a limiting magnitude of $r \lesssim19$~mag. Details of DLT40's GW counterpart search program are provided in \citet{valenti_discovery_2017} and \citet{yang_dlt40gw_2019}. The DLT40 galaxy catalog is drawn from the Gravitational Wave Galaxy Catalog \citep[GWGC;][]{White+11_GWGC} with further cuts on recessional velocity ($<$7000 km s$^{-1}$), declination ($<$ $+20^{\circ}$), absolute magnitude ($M_B < -18$~mag), and Milky Way extinction ($A_V < 0.5$~mag). Since LIGO/Virgo's Observing Run 2 (O2), DLT40 has built imaging templates for galaxies matching the above criteria out to $\sim$$100$~Mpc, beyond the original $\sim$$40$~Mpc, specifically for GW follow-up. For S251112cm, DLT40 observed 184 unique galaxies within the localization region. Several galaxies were visited multiple times.

\begin{figure}
\includegraphics[width=0.47\textwidth]{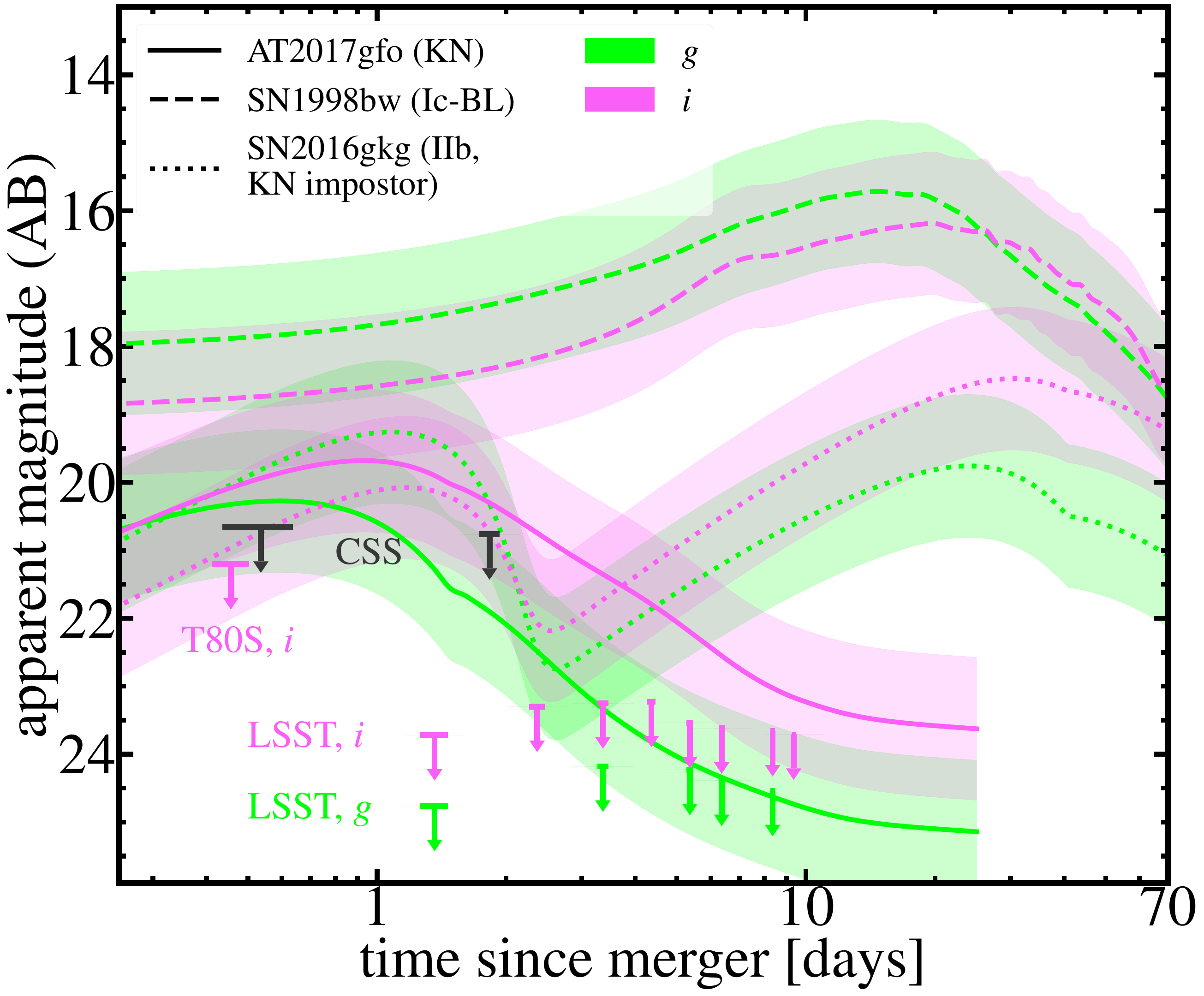}
    \caption{\textbf{Depths achieved by our CSS and T80-S observations, compared to a prototypical KN, SN\,Ic-BL, and SN\,IIb.} Models are generated from {\tt redback} following an AT\,2017gfo-like kilonova, SN\,1998bw (Ic-BL), and SN\,2016gkg (IIb; a class of SESNe but also noted KN impostors), scaled to the GW luminosity distance $93 \pm 27~\mathrm{Mpc}$ to S251112cm (Section~\ref{ssc:obs-search}). Translucent bands reflect the uncertainty on this distance. The KN model is truncated at 25 days, beyond which it loses reliability. Our CSS and T80-S observations would have likely found any KNe, SNe\,Ic-BL, or SNe\,IIb, in our observed fields. We also include median $g$- and $i$-band depths achieved by Rubin-LSST, as the greatest depths achieved by any search (\citealt{gcn42707, gcn43257}).}\label{fig:limmags}
\end{figure}

With CSS, we detect one new transient, AT\,2025adjd \citep{2025TNSTR4580....1S}. With T80-S and DLT40, we do not recover any new transients. To further assess our ability to have detected transients of interest, we compare the depths achieved by our search to relevant models. Figure~\ref{fig:limmags} shows the limiting magnitudes obtained with our CSS and T80-S wide-field search, compared to models for an AT2017gfo-like KN, a SN\,Ic-BL, and a SN\,IIb. We also include the median depths achieved by Rubin-LSST, as the greatest depths achieved by any search \citep{gcn42707, gcn43257}. We use \texttt{redback} \citep{sarin_redback_2024} to generate these models, scaling all to the GW luminosity distance $93 \pm 27~\mathrm{Mpc}$ of S251112cm.  To produce AT2017gfo-like lightcurves, we use the {\tt three\_component\_kilonova} model, with parameters from the homogeneous, three-component model in \citet{villar_combined_2017}. To compare to a SN\,Ic-BL, we show the prototypical SN\,1998bw, using the {\tt sn1998bw\_template} model \citep{vincenzi19}. For a double-peaked SN\,IIb, we adopt a {\tt shock\_cooling\_and\_arnett} model with SN\,2016gkg-like envelope mass $0.03~M_{\odot}$, envelope radius 43~$R_{\odot}$, explosion energy 10$^{51}$~erg, total ejecta mass 2~$M_{\odot}$, ejecta velocity 8100~km~s$^{-1}$ \citep{arcavi17_sn2016gkg}, and set all other parameters to defaults. The depth of our CSS observations suggests that we could have detected KNe, SNe\,Ic-BL, and SNe\,IIb in our observed fields at early times. Rubin-LSST could have likewise detected any of these transients.

These models provide a useful benchmark because SNe\,Ic-BL and IIb are classes of SESNe, which may be the SNe in which KNe are embedded in KNe-in-SNe \citep{metzger_fragmentation_2024, chen_gravitational_2025, lerner26_fragmentation}. Super-KNe may also resemble SNe\,Ic-BL or IIb \citep{siegel22_superKN}. The SN\,IIb also demonstrates their ability to act as KN impostors at early times. Indeed, we emphasize that SNe\,Ic-BL or IIb, or other SESNe, are not viable counterparts to SSM GW events on their own. Clear deviations from these models which indicate the presence of $r$-process material, \eg, deviations in color, would be required to establish a candidate as more than a standard CCSN/SESN (Section ~\ref{ssc:scoring-photom}). Association with a GRB would also provide evidence for the presence of a collapsar.

\subsection{Public Optical and High-Energy Observations}\label{ssc:obs-public-optical}

The broader community used a suite of optical telescopes to search for a counterpart. Observations were largely reported through GCN alerts and to the TNS. Over the first $\sim$$\mathrm{day}$, the Gravitational-wave Optical Transient Observer (GOTO; \citealt{gcn42658}), BlackGEM/MeerLICHT \citep{gcn42663}, and the Asteroid Terrestrial-impact Last Alert System (ATLAS) \citep{gcn42666} reported five transients within the localization region. Within the next 10 days, additional candidates and observations were reported with FTW, the Zwicky Transient Facility (ZTF), DECam, GRANDMA, Rubin-LSST, the Wide Field Survey Telescope (WFST), J-GEM, SVOM/VT, the Lulin Observatory Telescope (LOT), GECKO/KMTNet, and ATLAS \citep{gcn42674, gcn42677, gcn42682, gcn42691, gcn42698, gcn42699, gcn42707, gcn42722, gcn42725, gcn42777, gcn42825, gcn42867}. The largest influx of candidates came from observations with Rubin-LSST, conducted from 1.4 to 9.4 days post merger, and posted to the TNS and announced via GCN at 45 days post-merger (\citealt{gcn43257}). This Rubin-LSST search increased the number of optical candidates in our sample from 359 to 426.

Higher-energy (X-ray and gamma-ray) observatories were also employed in the search for a counterpart. In gamma-rays, the Fermi-GBM and all-sky Glowbug observed the localization region at the time of the initial LVK alert, but identified no candidates \citep{gcn42655, gcn42746}. In X-rays, the all-sky MAXI/GSC ($2-20~\mathrm{keV}$) found no potential counterparts within hours of the initial alert \citep{gcn42659}. The Swift X-Ray Telescope (Swift-XRT; $0.2 - 10~\mathrm{keV}$) searched the GW localization region over 0.1 to 1.4 days, identifying 7 previously-uncataloged and 13 previously-cataloged candidates \citep{gcn42681}. Concurrently, the Einstein Probe's Follow-up X-ray Telescope (EP-FXT; $0.5 - 4~\mathrm{keV}$) searched from 0.01 to 1.8 days, identifying 10 new X-ray candidates \citep{gcn42714}. These X-ray candidates are of interest given the recent increase in the number of fast X-ray transients (FXTs), detected mostly by EP since 2024. A selection of FXTs have been associated with GRBs (\eg, \citealt{levan24_EP240315a}) and optical counterparts (\eg, \citealt{gillanders24_240315a}), including CCSNe and potentially compact object mergers \citep{eyles-ferris25_EP250108a-SN2025kg, jonker26_EP250207b}. AGN flares induced by BBH mergers may also have X-ray counterparts \citep{Kimura2021}.

\subsection{Vetting of Candidates with \texttt{TROVE} \& Optical Follow-up}\label{ssc:obs-TROVE-motivated}

\begin{deluxetable*}
{lccccccc}
\tabletypesize{\footnotesize}
\tablecaption{Spectra of Candidates and Host Galaxies\label{tab:spectra}}
\tablewidth{5pt}
\tablehead{
\colhead{Name} & 
\colhead{Observation Date [UTC]} & 
\colhead{Source} & 
\colhead{Type} & 
\colhead{$z$} & 
\colhead{Ref.}}
\startdata
SN\,2025adgp & 2025-11-15 & SOAR/Goodman & Ia & 0.046$\pm$0.003 & 2 \\
SN\,2025adhf & 2025-11-17 & SOAR/Goodman & Ia & 0.098$\pm$0.003 & 1 \\
SN\,2025adhs & 2025-11-15 & SOAR/Goodman & II & 0.044$\pm$0.003 & 2 \\
AT\,2025adht & 2025-11-26 & Bok/B\&C & -- & 0.1848$\pm$0.0001 & This paper \\
SN\,2025adim & 2025-11-17 & SOAR/Goodman & Ia & 0.091$\pm$0.003 & 1 \\
SN\,2025adiw & 2025-11-17 & SOAR/Goodman & Ia & 0.165$\pm$0.002 & 1 \\
SN\,2025adiz & 2025-11-14 & MMT/Binospec & Ia & 0.132$\pm$0.002 & This paper \\
SN\,2025adjf & 2025-11-17 & SOAR/Goodman & Ia & 0.124$\pm$0.002 & This paper \\
AT\,2025adra & 2025-11-26 & Bok/B\&C & -- & 0.0784$\pm$0.0001 & This paper \\
\enddata
\tablecomments{Spectroscopic observations with Bok/B\&C, MMT/Binospec, and SOAR/Goodman, as described in Section~\ref{ssc:obs-TROVE-motivated}. Spectral types for transient spectra are noted along with spectroscopic redshifts.  Where we measured the host-galaxy redshift, we omit spectral type. References for discovery and classification of each candidate are (1) \citet{gcn42724} and (2) \citet{gcn42796}.}
\end{deluxetable*}

\begin{figure}[!t]
\includegraphics[width=0.47\textwidth]{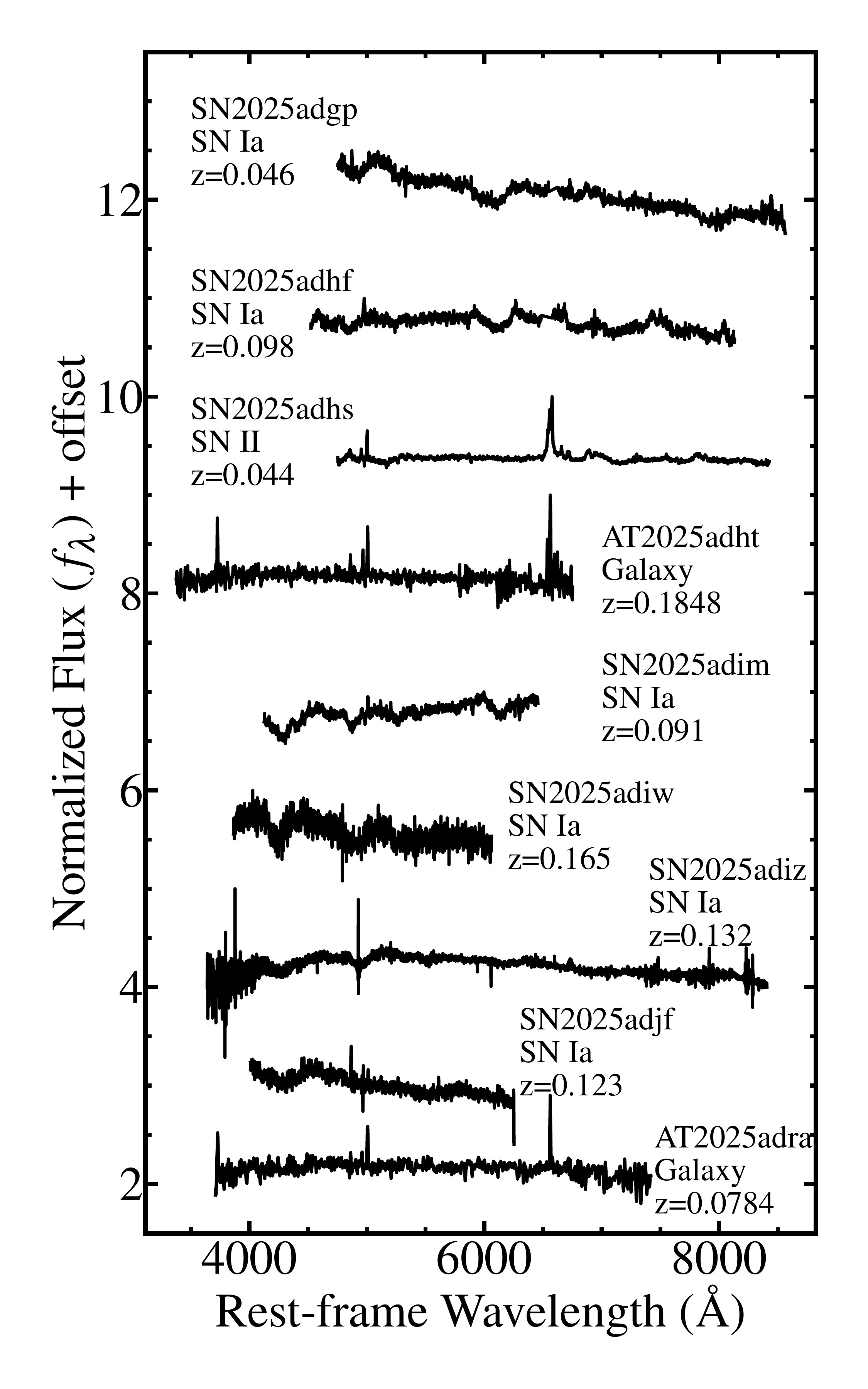}
    \caption{\textbf{Spectra of candidate counterparts or the putative hosts of candidate counterparts, with source classification and measured redshift.} Observations are compiled in Table~\ref{tab:spectra}. We obtained SOAR/Goodman spectra of five candidates (AT\,2025adgp, AT\,2025adhf, AT\,2025adhs, AT\,2025adim, AT\,2025adiw, and AT\,2025adjf}), MMT/Binospec spectra of AT\,2025adiz, and Bok/B\&C spectra of the host galaxies of AT\,2025adht and AT\,2025adra. All candidates resemble SNe\,Ia or II at redshifts $z$ too distant to correspond to S251112cm. All host galaxies similarly lie outside of the localization volume.\label{fig:spectra}
\end{figure}

Following the GW detection of S251112cm, we continuously ingested and vetted candidate counterparts from our searches and the community using the \texttt{TROVE} KN-vetting algorithm described in \franz25. Optical candidates were automatically ingested in near real-time directly from the TNS. For the small subset of optical candidates or photometry which were shared in GCNs but not to the TNS ($< 1\%$ of all candidates/photometry), we ingested these manually. We also manually ingested the 30 Swift-XRT and EP-FXT X-ray sources, reported in GCNs. We regularly queried the ATLAS forced photometry server \citep{tonry18_ATLAS, smith20_ATLAS, shingles21_ATLAS} to supplement candidates' lightcurves.

On 13 November 2025, at 31 hours post-merger, we released a GCN (\citealt{gcn42675}) identifying four candidates which received a high score with \texttt{TROVE}: AT\,2025adgp, AT\,2025adhf,  AT\,2025adhs, and AT\,2025adht. We included the caveats that some showed pre-GW detections with ATLAS and all  putative host galaxies' redshifts were only marginally consistent with the localization of S251112cm. Three of the four galaxies had photometric redshifts, with large uncertainties. Nonetheless, this list of candidates was then used by members of the community and ourselves in planning follow-up observations.

\subsubsection{Spectroscopic Follow-up}\label{sssc:obs-TROVE-motivated-spec}

We spectroscopically followed up on a total of six candidates using the Goodman High-Throughput Spectrograph \citep{Clemens04} on the 4.1 m Southern Astrophysical Research (SOAR) Telescope to classify these transients or their host galaxies (program PIs: (1) Bom and (2) Bommireddy). We show all follow-up spectra in Figure~\ref{fig:spectra} with the target, source classification, and measured redshift for each source. Our observations are compiled in Table~\ref{tab:spectra}. We processed all SOAR spectra following standard procedures with {\tt pypeit} \citep{pypeit:joss_arXiv,pypeit:joss_pub,pypeit:zenodo} using dome flat and arc lamp observations taken on the same night and instrumental configuration. Each spectrum was calibrated using a sensitivity function derived from a standard star spectrum obtained on the same night and corrected for telluric absorption. We systematically classified each spectrum using {\tt SNID-SAGE} \citep{snid_sage_2025} to derive their spectroscopic type and redshift. 

We obtained optical spectra of two of the four \texttt{TROVE}-identified candidates (AT\,2025adgp, AT\,2025adhs), and find them consistent with a SN\,Ia at $z = 0.046\pm0.003$ and SN\,II at $z = 0.044\pm0.003$, respectively; we released these redshifts and classifications by GCN shortly thereafter \citep{gcn42796} and reported all to TNS. We later obtained a spectrum of a third candidate (AT\,2025adhf), finding it consistent with a SN\,Ia at $z=0.098\pm0.003$, also released in a GCN \citep{gcn42724}. On the same night, we pursued three other candidates (AT\,2025adim, AT\,2025adiw, SN\,2025adjf) with notable scores and host galaxies consistent with the GW localization volume \citep{gcn42724}. We classify both AT\,2025adim and AT\,2025adiw as SNe\,Ia, at $z=0.091\pm0.003$ and $z=0.165\pm0.002$, respectively. We were initially unable to obtain a spectrum of AT\,2025adjf itself, so we obtained a spectrum of its host galaxy, measuring $z=0.124\pm0.002$. We later re-reduced its spectrum and were able to classify it as a SN Ia at the same redshift \citep{santos26_SN2025adjf}. Based on these  classifications and distances well outside the GW localization volume, we disqualify all six candidates as viable counterparts.

We also acquired a spectrum of one candidate (AT\,2025adiz) with Binospec on the 6.5 m MMT \citep{Fabricant2019}. Observation of this spectrum was triggered by the SAGUARO team using the {\tt PyMMT} package \citep{Shrestha2024}. As with other candidates, AT\,2025adiz had a noteworthy score and potential host with an uncertain photometric $z$. We find AT\,2025adiz consistent with a SN\,Ia at $z = 0.132 \pm 0.002$, outside the localization volume for S251112cm, disqualifying it. 

Lastly, we obtained spectra for the putative host galaxies of one candidate mentioned in our initial GCN (AT\,2025adht) and another notable candidate (AT\,2025adra) with the Boller \& Chivens (B\&C) Spectrograph on the 2.3 m Bok telescope. We find that the hosts of AT\,2025adht and AT\,2025adra both have emission line spectra and lie at $z = 0.1848 \pm 0.0001$ and $z = 0.0784 \pm 0.0001$, respectively, both well outside the localization volume of S251112cm. We thus disqualify AT\,2025adht and AT\,2025adra as well.

\subsubsection{Photometric Follow-up}\label{sssc:obs-TROVE-motivated-phot}

We also performed photometric follow-up on one X-ray source, S251112cm\_X78, initially reported by Swift-XRT \citep{gcn42681}, with SOAR-Goodman. Our follow-up was motivated by its association with a putative host galaxy with a photometric redshift $z = 0.02 \pm 0.01$ (Pan-STARRS1, \citealt{Beck+21}), within the localization volume of S251112cm. S251112cm\_X78 was previously uncataloged, though X-ray detections from $\sim 0.1 - 1.4~\mathrm{days}$~were not brighter than previous upper limits \citep{gcn42681}. We observed this field at 15.4 days post-merger for $6\times200$~s in the $g$- and $r$-bands, in relatively poor conditions. Processing these data using {\tt photpipe}, we searched for point-like sources in the 90\% containment localization of S251112cm\_X78. We detected no optical counterpart, down to limiting magnitudes of $g > 22.0$~and~$r > 21.87$.

AT\,2025adht, one of the candidates identified in our initial GCN, was followed up with the Maidanak Observatory on behalf of GRANDMA at 2.4 days \citep{gcn42698} and with the LOT at 9.2 and 10.2 days \citep{gcn42825}. 
After its initial detection with ZTF \citep{gcn42677}, this follow-up demonstrated that AT\,2025adht faded from 0.8 to 10.2 days at a rate slower than expected of a transient KN. \cite{gcn42825} interpret this evolution as consistent with an unrelated background SN near peak. Given our Bok/B\&C measurement of a redshift  $z = 0.1848 \pm 0.0001$ for the putative host galaxy, regardless of the nature of the transient, AT\,2025adht lies outside the GW localization volume.


\section{Scoring Candidate Counterparts to Subsolar Mass Gravitational-Wave Events} \label{sec:scoring}

\begin{figure*}[!ht]
    \centering
    \includegraphics[width=0.95\linewidth]{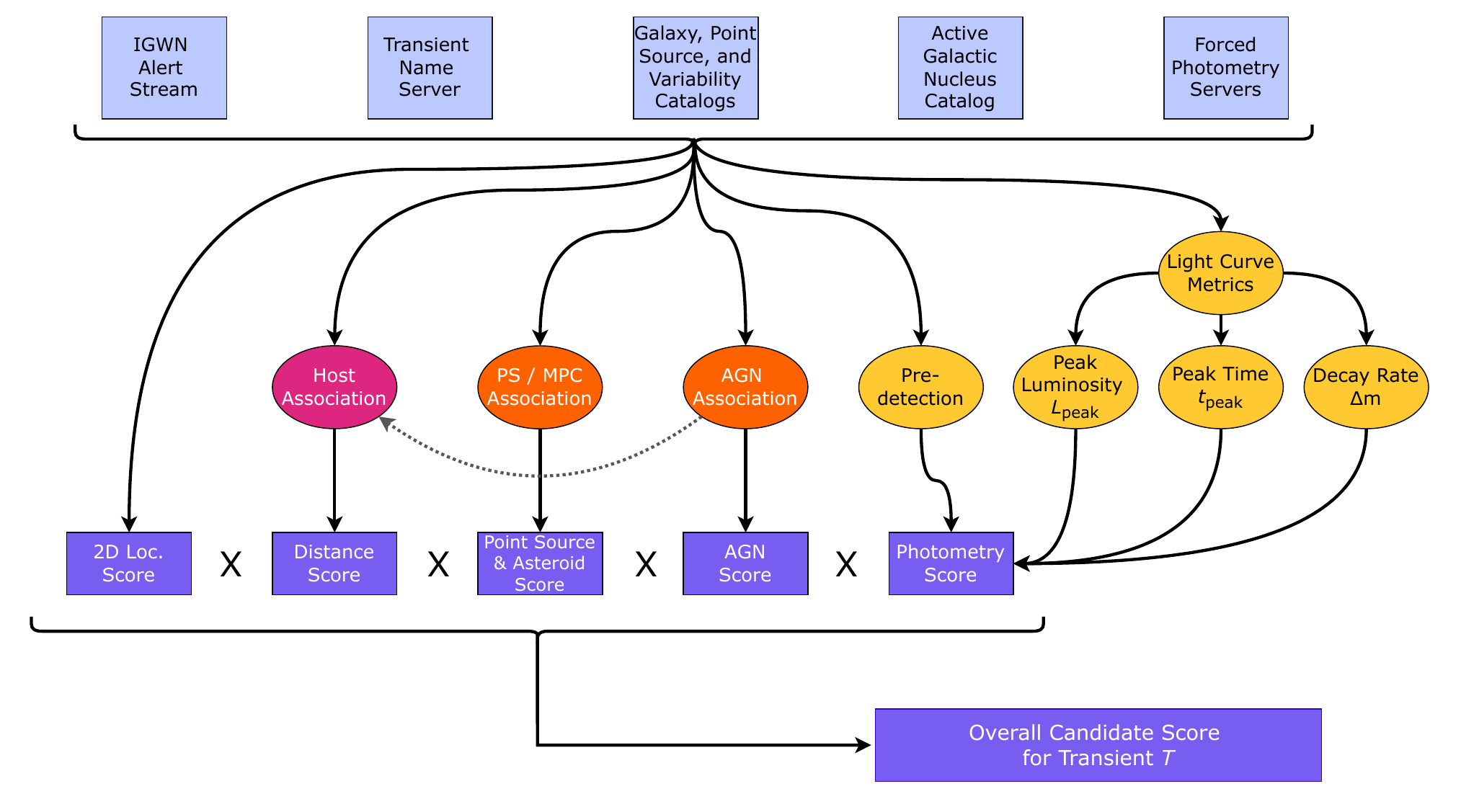}
    \caption{\textbf{Overall framework for scoring and ranking candidate counterparts to a SSM GW event.} We ingest GW alerts from the International Gravitational-Wave Observatory Network (IGWN) alert stream, transients and their photometry/classifications from the TNS, galaxies, point sources, variable objects, minor planets, and AGN from relevant catalogs, and forced photometry. For each candidate, GW alerts are used to compute the 2D localization subscore $S_{\mathrm{2D}}$ and in combination with galaxy catalogs to compute distance subscores $S_{\mathrm{dist}}$. Candidates' positions are compared to those of point sources, minor planets, and variable objects to compute $S_{\mathrm{PSMPC}}$. Association of candidates with AGN using an AGN catalog yields the AGN association subscore $S_{\mathrm{AGN}}$. If vetting for AGN flaring, the distance to any associated AGN is used in distance scoring, as represented with a dotted line. Photometry scores $S_{\mathrm{phot}}$ are computed by checking for predetections and comparing candidates' photometry to predictions for peak luminosity $L_{\mathrm{peak}}$, peak time $t_{\mathrm{peak}}$, and decay rate $\Delta m$, dependent on transient type $T$. Our framework introduces a dependence on the EM transient class $T$ in question (KNe, KNe-in-SNe, super-KNe, or AGN flares) for distance, point source/minor planet association, AGN association, and photometry scoring. Taking the product of the five subscores, we produce unique overall scores $S_T$ for each candidate for each class of EM transient $T$ (Equation~\ref{eqn:overallscore}).}
    \label{fig:scoring-diagram}
\end{figure*}

Our search for a counterpart to S251112cm outlined in the previous section leveraged our existing algorithm for vetting candidates in search of KN counterparts to GW events (\franz25). Here, we extend this algorithm and introduce metrics for vetting candidate counterparts to SSM events which scores candidates in comparison to multiple classes of transient. We develop a framework which searches for KNe, KNe-in-SNe, super-KNe, and BBH-induced AGN flaring. This framework can and should evolve as our knowledge of SSM mergers advances.

The score (between zero and one) for a candidate to potentially be transient $T$ that is also associated with a GW event is the product of five subscores: 

\begin{equation}\label{eqn:overallscore} 
    S_T = S_{\mathrm{2D}} \times S_{\mathrm{dist},T} \times S_{\mathrm{PSMPC},T} \times
    S_{\mathrm{AGN},T} \times
    S_{\mathrm{phot},T}~,
\end{equation}

\noindent where we use the subscript $T$~to indicate that the subscore is dependent on the transient type being vetted. We assign a subscore $S_{\mathrm{2D}}$ based on the candidate's 2D position in the GW localization region, calculating the total cumulative probability over all regions of the localization skymap at higher probability density (\franz25). $S_{\mathrm{dist},T}$ quantifies the overlap between the distance to the candidate (via direct spectroscopy or host galaxy) and the GW localization. We use $S_{\mathrm{PSMPC},T}$ to discard candidates associated with point sources or minor planets, and the newly-implemented $S_{\mathrm{AGN},T}$ favors or disfavors candidates associated with AGN. Finally, $S_{\mathrm{phot},T}$ is computed by comparing photometry to expectations for a given transient. Our overall framework is presented in Figure~\ref{fig:scoring-diagram}. 

In this framework, the score described in \franz25 would be $S_{\mathrm{KN}}$, with the caveat that \franz25~did not incorporate 2D AGN association into scoring.\footnote{However, both Gaia Data Release 3 (DR3; \citealt{gaiadr3}) and the Pan-STARRS 1 Source Types and Redshifts with Machine Learning catalog (PS1-STRM; \citealt{Beck+21}), used in \franz25~for point source association, contain AGN.} For known counterpart AT\,2017gfo to GW170817, we compute an overall score of $S_{\mathrm{KN}} = 0.54$ ($S_{\mathrm{KN}} = 0.56$)~if using the photometric (spectroscopic) $z$ of the host galaxy NGC 4993.
Here, we introduce the following new overall scores: $S_{\mathrm{KN\text{-}in\text{-}SN}}$, $S_{\mathrm{super\text{-}KN}}$, and $S_{\mathrm{AGN\text{-}flare}}$. Our 2D localization subscore $S_{\mathrm{2D}}$ is identical to that presented in \franz25. We describe the new $S_{\mathrm{AGN}}$, updates to $S_{\mathrm{PSMPC}}$ and $S_{\mathrm{dist}}$, and differences in $S_{\mathrm{phot}}$ across transients.

\subsection{AGN Association Score}\label{ssc:scoring-AGN}

We introduce a new subscore for 2D association with AGN. For most of the relevant transients (KNe, KNe-in-SNe, and super-KNe), we make the simplifying assumption that direct (a $\sim$few kpc) association with an AGN should disqualify the candidate as being the EM counterpart to the GW event. When vetting for BBH-induced AGN flaring, association with an AGN should instead favor the candidate. We thus define $S_{\mathrm{AGN},T}$, which takes on a value of zero or one dependent on the transient in question and association with an AGN or lack thereof. When vetting for KNe, KNe-in-SNe, or super-KNe (\ie, when $T =$ KN, KN-in-SN, or super-KN), a candidate is assigned $S_{\mathrm{AGN},T} = 0$ when associated with an AGN, disfavoring the candidate, and 1 when unassociated. When vetting for AGN flares ($T =$ AGN-flare), a candidate is assigned $S_{\mathrm{AGN},T} = 1$ when associated with an AGN, favoring the candidate, and 0 when unassociated.

We crossmatch candidates with AGN using version 8 of the Million Quasars (Milliquas) catalog \citep{flesch23_milliquasv8}. We consider a candidate associated with an AGN if it lies within 2$''$ ($\sim$$1~\mathrm{kpc}$ at $\sim$$93~\mathrm{Mpc}$) of said AGN, consistent with our criterion for point source association.

\subsection{Point Source/Minor Planet Association Score}\label{ssc:scoring-psmpc}

We define this subscore as the product of two score factors, $S_{\mathrm{PSMPC}} = S_{\mathrm{PS}} \times S_{\mathrm{MPC}}$, for point source (PS) and Minor Planet Center\footnote{\href{https://www.minorplanetcenter.net}{https://www.minorplanetcenter.net}} (MPC) object association. We set the score factor to zero for candidates that are confidently identified with known point sources or asteroids, and to one otherwise.

As in \franz25, we employ several catalogs for point source association.\footnote{The All-Sky Automated Survey for Supernovae Variable Star Catalog X (ASAS-SN X; \citealt{shappee_man_2014, christy_asas-sn_2023}), Gaia Data Release 3 (DR3; \citealt{gaiadr3}), and Pan-STARRS 1 Source Types and Redshifts with Machine learning catalog (PS1-STRM; \citealt{Beck+21}).} Here, we also include the ZTF Variable Stars catalog \citep{chen2020_ztfvarstar}. To exclude galaxies in PS1-STRM from our catalog of point sources, we ignore objects with a probability of being a galaxy (as computed in \citealt{Beck+21}) of $>$$0.7$. When vetting for BBH-induced AGN flaring, since AGN association should favor a candidate, association with a PS1-STRM object with probability of being a quasar $>$$0.7$ also does not disfavor the candidate. 

Our minor planet association largely follows that of \franz25, where we query the MPC and perform a two-body evolution around the Sun to identify asteroids within $10^{\circ}$ of a candidate at the time of its detection. For any matches, we then perform a full $n$-body simulation to find the precise location of the asteroid, and we consider candidates within 25$''$ of an asteroid to be associated with said asteroid. In this work, we impose an additional constraint: we consider any candidate with more than one $\geqslant$$5\sigma$ detection not to be an asteroid. This procedure aligns with our expectations for contamination from asteroids: the moving asteroid should be detected once at some position associated with a candidate, and then move on from that position.

\subsection{Distance Score}\label{ssc:scoring-distance}

We also update how we calculate the distance subscore. If the candidate has a known distance (\eg, a SN with some measured spectroscopic $z$) we select that distance. If no distance is known, we instead search for galaxies within 5$^{\prime}$ ($\sim$$150~\mathrm{kpc}$ at $\sim$$93~\mathrm{Mpc}$) and with probability of chance coincidence (\citealt{bloom_observed_2002}) $P_{\mathrm{cc}} \leqslant 0.15$, following \citet{rastinejad_systematic_2022}. For galaxies which pass this filtering, we then assign a potential host based on the provenance of their distance measurement, prioritizing (in descending order): redshift-independent measurements, spectroscopic $z$, and photometric $z$. If any galaxies possess a redshift-independent distance measurement, we compute the distance score $S_{\mathrm{dist}}$ as the maximum among all galaxies with redshift-independent measurements. If no such galaxy exists, we move on to spectroscopic $z$, and so on to photometric $z$. Finally, if no galaxies with $P_{\mathrm{cc}} \leqslant 0.15$ reside within 5$^{\prime}$, distance scoring is not factored in to the candidate's score, and $S_{\mathrm{dist}} = 1.0$. 

As in \franz25, we use a number of galaxy catalogs.\footnote{ Galaxy List for the Advanced Detector Era + (GLADE+; \citealt{dalya22_GLADEplus}), Gravitational Wave Galaxy Catalog (GWGC; \citealt{White+11_GWGC}), Heraklion Extragalactic Catalogue (HECATE; \citealt{Kovlakas+21_HECATE}), Legacy Survey Data Release 10 (LS DR10; \citealt{Zhou+23_LSDR10}), PS1-STRM \citep{Beck+21}, and Sloan Digital Sky Survey Data Release 12 (SDSS DR12) photo-$z$ catalog \citep{Alam+15}.} In this work, we replace the DESI EDR with Data Release 1 (DR1; \citealt{DESI26_DR1}) and incorporate the NASA/IPAC Extragalactic Database Local Volume Sample (NED-LVS; \citealt{cook2023_NEDLVS}). The completeness of galaxy catalogs at these nearby distances merits further exploration. However, \citet{tranin26_regalade} construct a set of catalogs similar to ours and find a completeness $>90$\% for galaxies contributing 50\% of the total $r$-band luminosity out to 360 Mpc, and completeness near 100\% out to the GW luminosity distance of S251112cm. This galaxy catalog completeness will be enhanced by upcoming surveys, and especially future data releases from DESI.

When vetting for BBH-induced AGN flaring, if a candidate is associated with an AGN, that AGN is taken to be the host and the distance to the AGN is used. To obtain uncertainties on the distance to the AGN, we follow Milliquas \citep{flesch23_milliquasv8} guidelines: for AGN with only a known photometric $z$, we assign an uncertainty $\Delta z = 0.1z~(0.01z)$ when $z$ is known to one (two) decimal places. When a spectroscopic $z$ is available, we set the uncertainty to a constant $\Delta z = 10^{-3}$.

\subsection{Photometry Score}\label{ssc:scoring-photom}

\begin{deluxetable*}{cc|cccc}
\tablecaption{Metrics for Computing Photometry Subscore}\label{tab:photom-metrics}
\tablehead{
\colhead{GW progenitor} &
\colhead{EM signature} & 
\colhead{pre-detections} & 
\colhead{$L_{\mathrm{peak}}$} &
\colhead{$t_{\mathrm{peak}}$} & 
\colhead{$\Delta m$} \\[-7pt]
\colhead{} &
\colhead{} & 
\colhead{allowed?} & 
\colhead{$[\mathrm{erg~s^{-1}}]$} &
\colhead{$[\mathrm{days}]$} & 
\colhead{$[\mathrm{mag~day^{-1}}]$}
} 
\startdata
BNS / NS+BH merger & KN & no & $< 10^{43}$ & $< 4$ & $>0.1$   \\
ssNS merger in $\sim$$1-10~M_{\odot}$ collapsar disk & KN-in-SN\tablenotemark{a}~$\approx$~SESN & $<1$~day & $[5\times10^{41},~10^{44}]$ & $< 35$ & $[-2.0, +0.1]$ \\
ssNS merger in $\sim$$10-50~M_{\odot}$ collapsar disk & super-KN\tablenotemark{b} & $< 1~\mathrm{day}$ & $[10^{41},~10^{43}]$ & $[10, 70]$ & - \\
BBH merger in AGN disk & AGN flare & yes & - & - & - \\
\enddata
\tablecomments{Photometry constraints (whether pre-GW detection is allowed, and allowed peak luminosities, rise times, and decay rates), for different classes of EM transient which could accompany a particular GW progenitor. Where left empty with `-', the metric is not usefully constrained by available observations or models.}
\tablenotetext{a}{\cite{metzger_fragmentation_2024} and \citet{chen_gravitational_2025}}
\vspace{-5pt}
\tablenotetext{b}{\cite{siegel22_superKN}}
\vspace{-15pt}
\end{deluxetable*}

Our photometric vetting scores candidate counterparts based on (1) the presence of any pre-GW detections, (2) the peak luminosity, (3) the epoch at which lightcurve peaks (\ie, rise time), and (4) the rate of evolution in their lightcurve (\ie, decay rate), for some class of transient $T$. We define the subscore for photometry:

\begin{equation}
    S_{\mathrm{phot},T} = S_{\mathrm{PD},T} \times S_{\mathrm{L},T} \times S_{\mathrm{RT},T} \times S_{\mathrm{DR},T} ~, 
\end{equation}

\noindent where the four score factors each take values of 1.0 if a condition is met and 0.1 if not. For example, for KNe, we require no predetections, a peak luminosity $L_{\mathrm{peak}} < 10^{43}~\mathrm{erg~s^{-1}}$, a rise time $t_{\mathrm{peak}} < 4~\mathrm{days}$, and decay rate $\Delta m > 0.1~\mathrm{mag~day^{-1}}$. A candidate which meets all of these criteria will have $S_{\mathrm{phot,KN}} = 1$; a candidate which meets none will have $S_{\mathrm{phot,KN}} = 10^{-4}$.

Our calculation of the maximum luminosity $L_{\mathrm{peak}}$ differs from that of \franz25. In \franz25, the maximum luminosity $\nu L_{\nu}$ is computed by converting the flux density $F_{\nu}$ at the brightest observed point in the candidate's lightcurve into the luminosity at the GW luminosity distance. Here, we instead use the same distance prioritization scheme as used in computing the distance score. The GW luminosity distance is used if and only if no host is identified. 

Most substantially, the metrics for computing score factors which multiply to generate the $S_{\mathrm{phot},T}$ different classes of transient. We summarize the different photometry metrics in Table~\ref{tab:photom-metrics}. These metrics also inform the time up to which we ingest new candidates. Based on predictions for super-KNe, the slowest-rising transient among those considered, we ingest no new candidates with initial detections later than 10 days post-merger. Super-KNe attain at least $\sim$50\% of their peak luminosity by this time, though they may continue to rise in brightness. Finally, metrics inform the time window over which we extract photometric quantities. For KNe, we extract rise times and decay rates considering photometry from only the first 25 days post-merger, following \citet{rastinejad_systematic_2022}. For KNe-in-SNe and super-KNe, we consider photometry up until 100 days, given the longer timescales of the transients. In all cases, we only use photometry detected at $\geqslant$$5\sigma$. 

In the KN-in-SN model (\citealt{metzger_fragmentation_2024, chen_gravitational_2025, lerner26_fragmentation}), accretion of the collapsar disk onto the central remnant BH occurs on timescales of seconds to minutes, while merger(s) of ssNSs occur over hours to $\lesssim 1~\mathrm{day}$. To account for this possible delay between the beginning of a transient SN and a GW signal, we conservatively allow for predetection of a transient, but no earlier than $1~\mathrm{day}$ pre-merger. We impose the same constraint for super-KNe.

We elaborate on luminosity, rise time, and decay rate metrics for KNe-in-SNe and super-KNe below. For BBH-induced AGN flares, we are unable to set useful constraints on luminosities, rise times, or decay rates; we discuss the challenges in doing so in Appendix~\ref{app:BBHAGN}.

\subsubsection{Kilonovae-within-Supernovae}\label{sssc:KN-in-SN}

We first identify the classes of SNe to which we will compare. While SNe\,Ic-BL are frequently coincident with lGRBs and thought to arise at least in part from collapsars \citep{woosley06_lGRBSN}, these are not the only SESNe which might be associated with collapsars \citep{woosley93, macfaydenwoosley99, dainotti22_GRBSNe}, so we broaden our horizons and consider SNe\,Ib, Ic, Ic-BL, and IIb. These SESNe may then be modified by the presence of the high-opacity $r$-process material from ssNS mergers in the collapsar disk or energy injection from a hierarchically formed millisecond magnetar. However, searches for $r$-process signatures in SNe Ic-BL \citep{anand_rproc_collapsars_SNIcBL_2024} and GRB-SNe \citep{rastinejad24} have thus far not confirmed the presence of $r$-process elements, and modeling work is ongoing (\eg, \citealt{barnes_2022, giacomo25}). We therefore simplify our expectations for an EM signature and simply vet for SESNe, but add some buffer to our expected peak luminosities, rise times, and decay rates to account for deviations from standard SESNe. Indeed, we expect KNe-in-SNe to deviate from standard SESNe, and only clear deviations may establish a candidate as more than a normal SESN.

We explore a sample of SESNe to identify broad photometric behavior.  \citet{rodriguez23_SESNe} examine a sample of 191 SESNe, expanding on the samples in \citet{lyman16_SESNe} and \citet{prentice19_SESNe}. \citet{taddia19_IcBL} and  \citet{srinivasaragavan24_IcBL} focus specifically on SNe\,Ic-BL. Broadly, we find that SESNe peak at bolometric luminosities in the range 
$L_{\mathrm{bol,peak}} \sim 7 \times 10^{41}$ to $5\times 10^{43}~\mathrm{erg~s^{-1}}$, between $\sim$$3$ and $\sim$$33$ days from explosion. SESNe display post-peak decay rates of approximately $0.01$ to at steepest $0.1~\mathrm{mag~day^{-1}}$. Finally, they brighten by at most approximately $-0.3~\mathrm{mag~day^{-1}}$ before peak.

We use these samples to constrain the luminosity, rise time, and decay rate that we expect of KNe-in-SNe. We note that we have compiled here samples of $L_{\mathrm{bol,peak}}$, the peak bolometric luminosity, but we estimate the peak luminosity with \texttt{TROVE} as $L_{\mathrm{peak}} = \nu L_{\nu}$, where $\nu$ is the effective frequency of the filter at the brightest observed point, and $L_{\mathrm{bol}} > \nu L_\nu$ by definition. Allowing some buffer on the lower bound due to differences between $L_{\mathrm{bol}}$ and $\nu L_\nu$, and on both lower and upper bounds to account for deviations from SESNe, we impose that $L_{\mathrm{peak}}$ must lie between $5\times10^{41}$ and $ 10^{44}~\mathrm{erg~s^{-1}}$. Allowing for a slightly earlier or later peak than in typical SESNe, we adopt that the rise time $t_\mathrm{peak} < 35~\mathrm{days}$. Finally, we apply that the decay rate $\Delta m < 0.1~\mathrm{mag~day^{-1}}$. We limit how rapidly a candidate may fade (\ie, apply an upper limit on $\Delta m$) in vetting for KNe-in-SNe to contrast with vetting for KNe, which requires that the candidate fade sufficiently rapidly. The relevant SESNe (with the exception of the shock-cooling phase in some SNe\,IIb, which may mimic the rapid decline in KNe) evolve more slowly than isolated KNe, and the KN embedded in the SN might further slow the decline by extending the photospheric phase \citep{barnes_2022}. However, to ensure that not all brightening is allowed, we also impose a lower limit on $\Delta m$. The compiled samples generally brighten by at most $\sim$$-0.3~\mathrm{mag~day^{-1}}$ in the 10 days pre-peak. We invoke for the sake of example the  exceptionally fast-rising SN\,(II-P) 2018lab, which rises by $\sim$$-1 ~{\rm to}~-2~\mathrm{mag~day^{-1}}$ in the first day post-explosion \citep{pearson_circumstellar_2023}. SNe\,II-P are not included in our list of SNe in which a KN might be embedded. We therefore finally impose $ -2~\mathrm{mag~day^{-1}} < \Delta m < +0.1~\mathrm{mag~day^{-1}}$. This lower limit is conservative and satisfied by all but the most rapidly brightening transients.

These metrics allow us to pick out SESNe spatially and temporally associated with a SSM GW event based on $S_{\mathrm{KN\text{-}in\text{-}SN}}$ scores. However, a high score alone is not sufficient to establish the presence of a KN embedded in a SESN; high scores motivate follow-up. As modeled in \citet{barnes_2022}, a clear signature of $r$-process material from a KN-in-SN would be a near-infrared (NIR) excess at late times, manifesting with colors $r-J$, $r-H$, and $r-K$ in excess of $\approx$$1-2$, especially after $\sim$$75~\mathrm{days}$. $r$-process material might also shift rise times and extend the photospheric phase of the SN. Comparisons to such models, as conducted in \citet{rastinejad24} and \citet{anand_rproc_collapsars_SNIcBL_2024}, are a promising avenue for assessing whether $r$-process material is present. These comparisons would benefit from sustained, multi-band optical/IR monitoring up to late times $\sim$$75-100~\mathrm{days}$, as will be supported by LSST.

\subsubsection{Super-Kilonovae}\label{sssc:superKN}

As \cite{siegel22_superKN} present the only detailed exploration of optical/IR signatures of collapsars with $r$-process material enrichment in the relevant mass regime, we rely on their predictions. \citet{siegel22_superKN} explore models spanning total ejecta masses of $8.6 - 60.0~M_{\odot}$, with differing proportions of $r$-process and non-$r$-process (including ${}^{56}$Ni) material. Their bolometric lightcurves peak at $\sim$$15 - 60~\mathrm{days}$ and luminosities $\sim$$5 \times 10^{41} - 5 \times 10^{42}~\mathrm{erg~s^{-1}}$. We thus conservatively impose constraints of $10^{41}~\mathrm{erg~s^{-1}} < L_{\mathrm{peak}} < 10^{43}~\mathrm{erg~s^{-1}}$ and $10 ~\mathrm{days} < t_{\mathrm{peak}} < 70~\mathrm{days}$. 

Because the different presented models brighten or fade by factors of $\sim$$10 - 100 \times$ over several weeks depending on the model, we cannot impose any useful constraints on $\Delta m$. These large variations in photometric behavior across models arise from differences in total ejecta mass and the relative proportions of $r$-process and ${}^{56}$Ni ejecta, which yield different radioactive heating rates driving the time evolution of the transient. \cite{siegel22_superKN} note that detailed nucleosynthesis calculations could better constrain the composition of the ejecta, and thus this photometric evolution.

These imposed metrics are broad, but select slow-rising candidates at predicted luminosities. Nonetheless, as with KNe-in-SNe, a high score is not sufficient to argue that a candidate is a super-KN, and follow-up is required. \cite{siegel22_superKN} compare their predicted bolometric luminosities to a selection of SNe\,Ia (SN\,2011fe), Ic-BL (SN\,2002ap), II-P (SN\,2013ab), and electron-capture SNe (ECSNe; SN\,2018zd). They find that super-KNe do not attain the peak luminosities of SN\,2011fe (Ia) nor SN\,2018zd (ECSNe), but may masquerade as SNe\,Ic-BL or II-P at early times $\lesssim$$50$~days. Such SNe might thus receive a high super-KN score. Spectra could be a key differentiator, as the spectra of super-KNe are predicted to be redder and display broader absorption features than in SNe due to the presence of faster ($v \sim 0.1c$) outflows and higher-opacity material. Though SNe\,Ic-BL show similar broad absorption features \citep{modjaz16_SNIc-BL_spectra, finneran25_SNIc-BL_spectra}, these features in super-KNe should emerge at redder wavelengths $\gtrsim$$10,000$~\AA. We concur that observations of these red spectra and broad absorption features, near peak luminosity ($\sim$$10-70~\mathrm{days}$), are a minimum for identifying a candidate as a potential super-KN.


\section{Scoring All Candidate Counterparts to S251112\lowercase{cm}}\label{sec:candscores}

With our vetting framework, we assign scores to all 456 transients in the localization region of S251112cm with initial detections up to 10 days following the GW alert. We compute final scores using all available information at 100 days. Tables~\ref{tab:scores_allcands} and \ref{tab:photom_allcands} (Appendix~\ref{app:scores-all}) contain detailed scores, whether the candidate is detected pre-GW signal, and estimates for luminosities, rise times and decay rates which inform photometry scores, for all candidates. We examine broad trends in scores and our ability to distinguish KNe, KNe-in-SNe, and super-KNe in Section~\ref{ssc:candscores-scores}, and the time evolution of scores in Section~\ref{ssc:candscores-time}. We highlight candidates of particular interest in Section~\ref{ssc:candscores-interesting}. Because only two candidates are associated with AGN, we primarily focus on KNe, KNe-in-SNe, and super-KNe. Throughout, we emphasize that our framework assigning high scores to candidates with limited information is by design, as these are some of the candidates which might have benefited most from follow-up. Indeed, the aim of \texttt{TROVE} is to motivate the follow-up which is necessary to identify a true counterpart.

\subsection{Scoring All Candidate Counterparts}\label{ssc:candscores-scores}

In all, 174 of 456 candidates receive an overall score for at least one transient class of $\geqslant 0.01$. These include 131 candidates for KNe, 149 for KNe-in-SNe, 120 for super-KNe, and 1 for AGN flaring. For some candidates, the overall score is the same across transient classes, especially when photometry is minimal or a distance is not known. Indeed, distance and photometry are the most discriminating factors. By construction, most candidates have 2D localization scores $S_{\mathrm{2D}} \geqslant 0.01$, with a few exceptions for candidates identified with Rubin-LSST. We find that 20 candidates are associated with point sources, two with minor planets, and six with AGN. These small numbers reflect that many teams already discard minor planets or candidates associated with point sources before reporting them in GCNs or to the TNS. 

Many candidates are disqualified due to a potential host galaxy or direct distance measurement that lies outside the GW localization volume. In total, over half (294) of all candidates receive distance scores $S_{\mathrm{dist}} \leqslant 0.10$, and 194 further below with $S_{\mathrm{dist}} \leqslant 0.01$. Only 27 candidates have neither a measured redshift nor a known potential host galaxy in our galaxy catalogs, such that distance does not factor into their overall score. The fact that a large number of candidates are disfavored based on existing information is consistent with other works \citep{kilpatrick_gravity_2021, rastinejad_systematic_2022}. Spectroscopic follow-up provides refined distance estimates (and classifications) for several candidates: 11 are directly classified as SNe confidently outside the localization volume, and two candidates' host galaxies are also found to lie outside the localization volume. 

\begin{figure}
\includegraphics[width=0.45\textwidth]{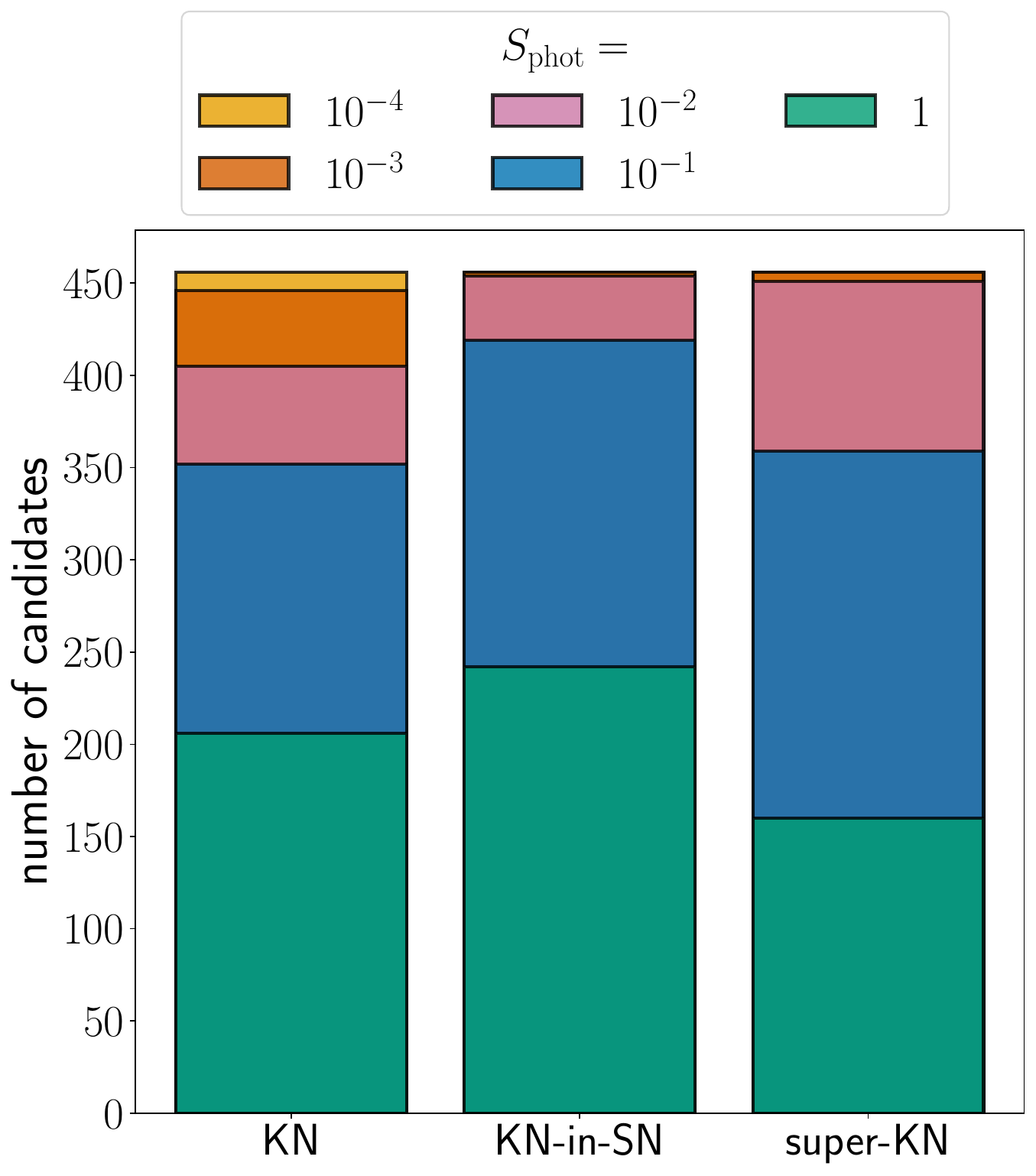}
    \caption{\textbf{Distributions of KN, KN-in-SN, and super-KN photometry subscores}. Relative distributions of photometry subscores reflect the constraining power of the photometry metrics imposed on candidates (Table~\ref{tab:photom-metrics}).}\label{fig:cands-photscores}
\end{figure}

In photometry scoring, we impose the constraints defined in Table~\ref{tab:photom-metrics} to further vet candidates and differentiate between potential transient types. Figure~\ref{fig:cands-photscores} shows the distribution of photometry scores ($S_{\mathrm{phot}} = $ 1.0, 0.1, 0.01, 0.001, or 0.0001) for each transient type. Overall, 250 candidates receive $S_{\mathrm{phot},\mathrm{KN}} \leqslant 0.10$, while 214 receive $S_{\mathrm{phot},\mathrm{KN\text{-}in\text{-}SN}} \leqslant 0.10$, and 296 receive $S_{\mathrm{phot},\mathrm{super\text{-}KN}} \leqslant 0.10$. These tallies are not mutually exclusive: 128 candidates have photometry scores of 1.0 for all transient types (\ie, are not constrained by photometry), while 34 have photometry scores of 0.1 (weakly constrained by photometry) for all transient types. The relative numbers of candidates in each tier of $S_{\mathrm{phot}}$ demonstrate the constraining power of each transient class' photometry metrics. We observe the following behavior from differing constraints: 

\begin{itemize}
    \item \textit{\textbf{Pre-detections:}} 36 candidates are considered predetected by our KN metrics. Since we allow that KNe-in-SNe and super-KNe may be predetected up to 1 day pre-merger, only 26 are predetected by KN-in-SN or super-KN metrics.
    
    \item \textit{\textbf{Luminosities:}} 185 candidates with $L_{\mathrm{peak}} > 10^{43}~\mathrm{erg~s^{-1}}$ are disfavored as KNe or super-KNe. In comparison, only 95 candidates are disfavored as KNe-in-SNe due to the more permissive bounds on their luminosities.

    \item \textit{\textbf{Rise times:}} 151 candidates have an inferred rise time shorter than 10 days, disfavoring them as super-KNe. Only 76 are disfavored as KNe due to a rise time longer than 4 days, and just 31 are disfavored as KNe-in-SNe due to a rise time longer than 35 days.

    \item \textit{\textbf{Decay rates:}} 118 candidates are disfavored as KNe due to an inferred $\Delta m < 0.1~\mathrm{mag~day^{-1}}$ (fading too slowly or brightening). 93 candidates are disfavored as KNe-in-SNe due to an inferred $\Delta m > 0.1~\mathrm{mag~day^{-1}}$ (fading too rapidly), and just 13 have $\Delta m < -2.0~\mathrm{mag~day^{-1}}$.
\end{itemize}  

Summarizing these trends, we find that $S_{\mathrm{phot},\mathrm{KN}}$ scores are largely shaped by luminosities, followed by decay rates, and then rise times. $S_{\mathrm{phot},\mathrm{KN\text{-}in\text{-}SN}}$ scores are shaped by decay rate estimates, some candidates disfavored due to their luminosities, and a comparatively negligible number due to their rise times. $S_{\mathrm{phot},\mathrm{super\text{-}KN}}$ scores are most strongly impacted by rise time estimates, with some contribution from luminosity estimates. Our imposed metrics result in super-KN scores consistently being the same or smaller than KN or KN-in-SN scores for a given candidate. We ascribe these small scores to the significant number of candidates with rise times $< 10~\mathrm{days}$, which reflects observational strategies and the concentration of publicly available photometry at early times, as addressed in the next section.

Despite the different constraining powers of our metrics, overall ranges in scores are comparable for KNe, KNe-in-SNe, and super-KNe. However, we reiterate that a high score alone is not sufficient to establish a candidate as a KN, KN-in-SN, or super-KN, and selecting the highest score among the three is not a tested form of classification. Examining candidates' scores as a function of time can provide further insight.

\subsection{Time Evolution of Candidate Scores}\label{ssc:candscores-time}

\begin{figure}
\includegraphics[width=0.47\textwidth]{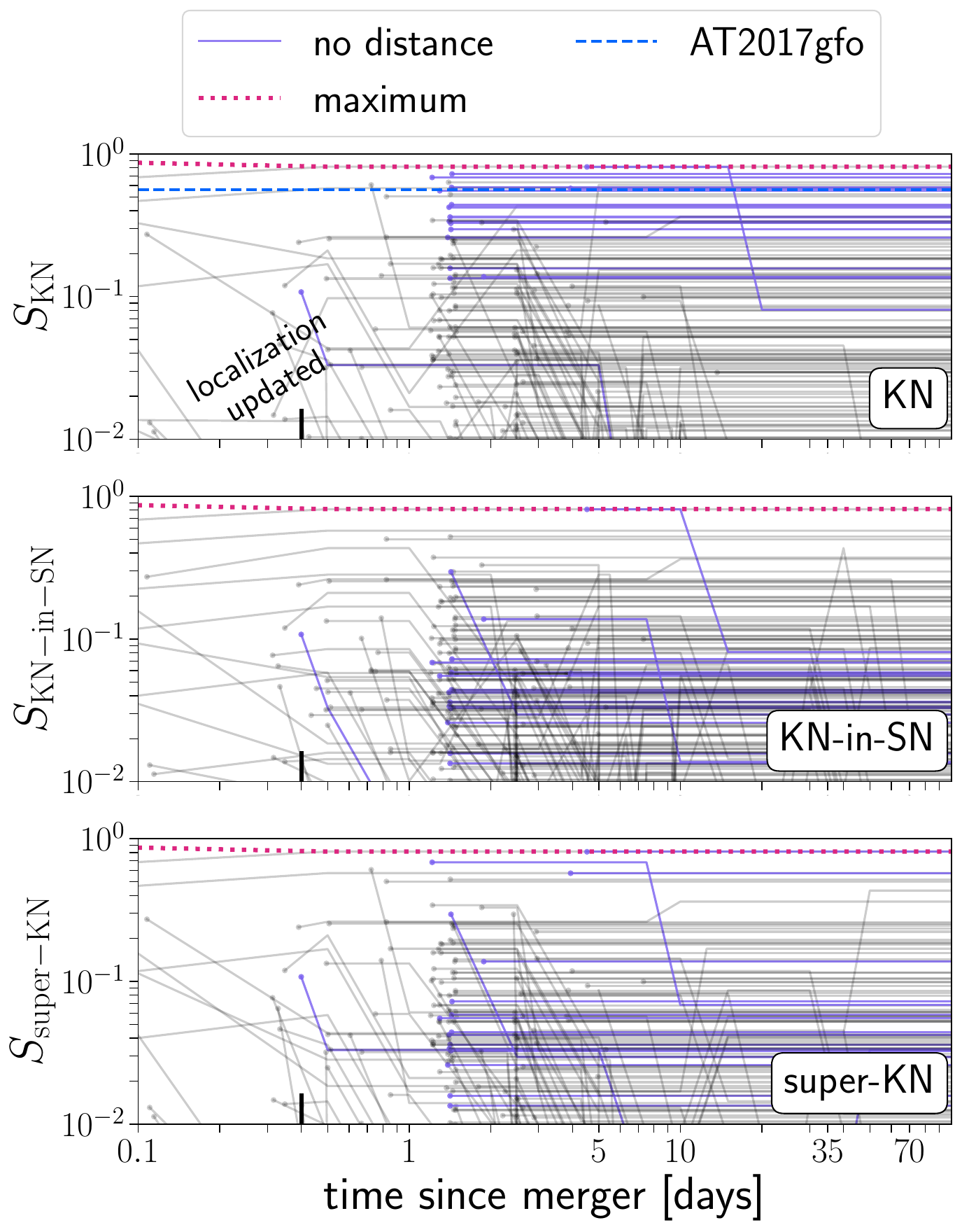}
    \caption{\textbf{Overall scores as a function of time for all candidate optical counterparts to S251112cm.} Each trace tracks the score for one of 426 candidates. Traces begin (as marked with a dot) at the time of first $\geqslant$$5\sigma$ optical detection. Candidates with no distance score due to a lack of direct distance measurement or viable host galaxy are highlighted in purple. The localization region of S251112cm is updated at 0.4 days, but most candidates are discovered after this time. Several candidates are discovered at $\sim$$1.4$ days with DECam \citep{gcn42691} and at $\sim$$2.4$ days with Rubin-LSST \citep{gcn43257}. Candidates for which scores quickly drop are those targeted by spectroscopic follow-up, driving their distance score to $\sim$$0$ when the distance to the candidate or its host is found to lie outside the localization volume. We highlight the final score $S_{\mathrm{KN}} = 0.56$~of AT2017gfo, known KN counterpart to the BNS merger GW170817. The highest score across all candidates at any given time, for a given transient type, is highlighted with a dotted magenta line. This maximum need not correspond to the same candidate at all times (Section~\ref{ssc:candscores-time}).}\label{fig:all-cands-scores-over-time}
\end{figure}

\begin{figure}
\includegraphics[width=0.47\textwidth]{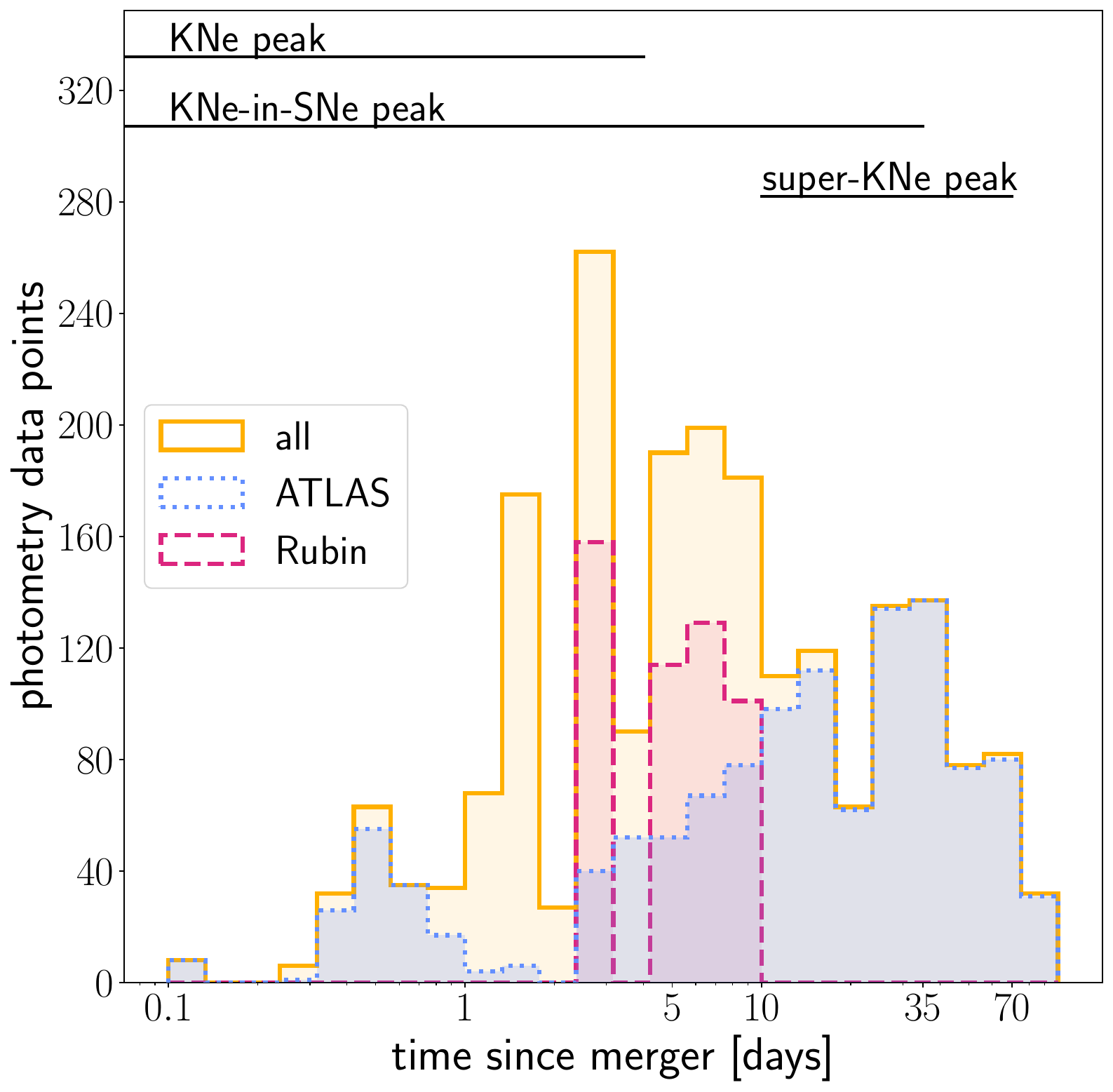}
    \caption{\textbf{Time distribution of photometry data points acquired for all candidates.} We distinguish ATLAS forced photometry (blue), Rubin-LSST photometry (magenta), and the sum of all photometry (yellow). Horizontal lines show the time range over which we expect KNe (up to 4 days), KNe-in-SNe (up to 35 days), and super-KNe ($10-70$~days) to peak. DECam photometry \citep{gcn42691} dominates at 1.4 days. Rubin-LSST photometry \citep{gcn43257} dominates at 2.4, 4.4, 6.4, 7.4, and 9.4 days. Beyond 10 days, when super-KNe are predicted to peak, most teams stopped acquiring photometry.}\label{fig:allphot-hist}
\end{figure}

Figure~\ref{fig:all-cands-scores-over-time} shows the time evolution of overall KN, KN-in-SN, and super-KN scores for all 426 optical candidates. We omit Swift-XRT and EP-FXT X-ray sources as these are not reliably detected in optical photometry.\footnote{Seven Swift-XRT sources are marginally ($\sim$$5\sigma$) detected with ATLAS, but not significantly brighter than limiting magnitudes at adjacent epochs, so these detections may be noise. We claim no optical counterparts to any X-ray sources.} We highlight candidates which have no distance score (purple, solid lines). As a benchmark, we highlight the final score $S_{\mathrm{KN}} = 0.56$ of AT2017gfo, known KN counterpart to BNS merger GW170817 (blue, dashed). Finally, we highlight the highest score (magenta, dotted line), which need not correspond to the same candidate at all times.  

Improved GW localization between the LVK's initial low-latency analysis and refined offline analysis, generally released within just a few days of the initial, can change 2D localization and distance scores for all candidates. For S251112cm, the refined offline analysis was released at 0.4 days. The updated skymap introduces changes in $S_{\mathrm{2D}}$ on average $\pm 0.2$ (at most $+0.5$ or $-0.4$), which propagate to largely negligible differences $\sim$$\pm0.01$ in overall scores. Moreover, most candidates are discovered after the release of this refined localization. Spectroscopic follow-up and classification leads twelve candidates' scores to drop to $\sim$$0$ once these establish the candidates as residing outside the GW localization volume or yield a classification (in most cases, SN\,Ia) inconsistent with predictions. All remaining evolution in scores is then a consequence of acquiring more photometry such that luminosity, rise time, and decay rate estimates evolve. Given the different timescales for the transients of interest, there is a natural point at which scores stop evolving. $S_{\mathrm{KN}}$ is frozen after 25 days as we do not include photometry beyond this time when estimating photometric quantities, following \citet{rastinejad_systematic_2022}, while $S_{\mathrm{KN\text{-}in\text{-}SN}}$ and $S_{\mathrm{super\text{-}KN}}$ may continue to evolve as we fit photometry up to 100 days.  

While the highest-scoring candidate (AT\,2025adtm, with $S_{\mathrm{KN}} = S_{\mathrm{KN\text{-}in\text{-}SN}} = S_{\mathrm{super\text{-}KN}} = 0.81$) at all times has a distance measurement, several candidates retain high scores due to the lack of a distance measurement (highlighted in purple in Figure~\ref{fig:all-cands-scores-over-time}) or limited photometry. In some cases, in the absence of a distance measurement, a single photometric detection yields $L_{\mathrm{peak}} < 10^{40}~\mathrm{erg~s^{-1}}$, too faint for either KNe-in-SNe or super-KNe at the GW luminosity distance. This is the case for AT\,2025adnj, AT\,2025adnk, AT\,2025adnm, AT\,2025adoa, AT\,2025adoe, AT\,2025adoh, AT\,2025adov, AT\,2025adpc, AT\,2025adpv, AT\,2025adpx, and AT\,2025aebg, which retain high $S_{\mathrm{KN}} \sim 0.20 - 0.70$ but low $S_{\mathrm{KN\text{-}in\text{-}SN}}$ and $S_{\mathrm{super\text{-}KN}} \sim 0.02 - 0.07$ due to this low peak luminosity estimate based on a single data point. More photometry for these candidates might have revealed brightening or fading. Candidates with distance measures broadly have lower scores, but many in addition to AT\,2025adtm (AT\,2025admn, AT\,2025admt, AT\,2025admu, AT\,2025adnf, AT\,2025adnp, AT\,2025adof, AT\,2025adrc, and AT\,2025adtn) nonetheless retain overall scores $S_{\mathrm{KN}} = S_{\mathrm{KN\text{-}in\text{-}SN}} = S_{\mathrm{super\text{-}KN}} \sim 0.10 - 0.60$ at all times when only one non-constraining photometry detection is available.   

Indeed, Figure~\ref{fig:all-cands-scores-over-time} demonstrates that scores do not evolve with time for a majority of candidates. This lack of evolution is due to a dearth of photometry. Notably, 192 candidates have only a single reported photometry data point. Excluding forced photometry obtained with ATLAS increases this number to 276. Photometry is also concentrated at early times. Figure~\ref{fig:allphot-hist} shows the time distribution of all photometry data, highlighting the contributions from Rubin-LSST photometry and ATLAS forced photometry. Initial post-GW detections are dominated by DECam (123 candidates), ATLAS (100), WFST (73), and Rubin-LSST (48), with additional contributions from GOTO-N and GOTO-S (30), PS1 (20), PS2 (15), ZTF (12), and BlackGEM (5). Taking the ensemble of all post-GW photometry, half is acquired prior to 4.5 days. For the 192 candidates with only a single non-ATLAS data point, half of all photometry is obtained much earlier, prior to 1.5 days. Beyond 10 days, at which times super-KNe are expected to peak, ATLAS is the only survey with publicly accessible data in the localization region. We thus caution that the observed tendency towards shorter rise times and the ensuing impact on scores, especially $S_{\mathrm{super\text{-}KN}}$, likely reflects the dearth of publicly available survey data outside of initial GCN and TNS discovery reports. 

In contrast, candidates with more sustained photometry show dynamic scores. Rubin-LSST photometry introduces these dynamics for several candidates. Longer-term ($\sim$weeks to months) and deeper photometry for candidates, as will be supported by regular operations of the LSST, will be critical to constraining transients which evolve more slowly. We highlight a sample of interesting candidates with dynamic and noteworthy scores which might have benefited from this longer-term monitoring in the next section.

\subsection{Candidates of Interest}\label{ssc:candscores-interesting}

\begin{deluxetable*}{ccccc}
\tablecaption{Candidates of Interest}\label{tab:interestingcands}
\tablehead{\colhead{candidate} & \colhead{ $S_{\mathrm{KN}}$ } & \colhead{$S_{\mathrm{KN\text{-}in\text{-}SN}}$} &  \colhead{$S_{\mathrm{super\text{-}KN}}$} & \colhead{$S_{\mathrm{AGN\text{-}flare}}$}}
\startdata
AT\,2025adin & \textbf{0.61} & 0.01 & 0.06 & 0.00 \\
AT\,2025admd & 0.00 & \textbf{0.13} & 0.00 & 0.00 \\
AT\,2025ailp & \textbf{0.02} & 0.00 & 0.00 & 0.00 \\
AT\,2025ails & 0.00 & \textbf{0.11} & 0.01 & 0.00 \\
S251112cm\_X3 & 0.00 & 0.00 & 0.00 & \textbf{0.13} \\
\enddata
\tablecomments{Candidate counterparts of interest, described in Section~\ref{ssc:candscores-interesting}. For each candidate, the highest score in the row is bolded. Scores of 0.00 correspond to scores of $<$$0.005$, which are rounded down to 0.00 given our trust in the precision of our scores.}
\end{deluxetable*}

For all 456 candidates, as available, we visually inspect their potential host galaxies, lightcurves, the time evolution of scores, and final scores. We perform this visual inspection post-facto to assess how our vetting performs and highlight candidates of interest; during a follow-up campaign, the aim of \texttt{TROVE} is to assist in automating this process. We parse our sample for candidates which we define as interesting for the purposes of this discussion---those with an overall score $\geqslant 0.01$ for at least one transient type and some combination of the following criteria: 

\begin{enumerate}
    \item A distance (to the candidate itself or to its likely host) which lies within or overlaps substantially with the localization volume of S251112cm ($S_{\mathrm{dist}} \gtrsim 0.1$), 
    \item Sufficient photometry for KN, KN-in-SN, and super-KN scores to differ, and/or, 
    \item Sufficient photometry for KN, KN-in-SN, and super-KN scores to evolve in some notable way with time.
\end{enumerate}

We highlight four candidates which meet these criteria in Table~\ref{tab:interestingcands}, receiving a notable score for $S_{\mathrm{KN}}$ or $S_{\mathrm{KN\text{-}in\text{-}SN}}$ and show interesting behavior in their lightcurves or scores with time. No candidates reliably receive a $S_{\mathrm{super\text{-}KN}}$ score greater than $S_{\mathrm{KN}}$ or $S_{\mathrm{KN\text{-}in\text{-}SN}}$. Unsurprisingly, all four candidates benefit from Rubin $g$- and $i$-band photometry. We discuss broad characteristics of these four candidates here and provide details, lightcurves, and scores with time for each candidate in Appendix~\ref{app:interest}. We also include in Table~\ref{tab:interestingcands} the X-ray candidate S251112cm\_X3, the only candidate to receive $S_{\mathrm{AGN}\text{-}\mathrm{flare}} > 0.01$ (Appendix~\ref{app:interest}). Finally, we also include the SN\,IIb of recent interest SN\,2025adtq \citep{hall26_S251112cm} in Appendix~\ref{app:interest}. 

Two candidates (AT\,2025adin and AT\,2025ailp) fare best as KNe. 2D scores range from $S_{\mathrm{2D}} = 0.06$ to $0.70$, and distance scores from $S_{\mathrm{dist}} = 0.23$ to $0.87$. AT\,2025adin and AT\,2025ailp receive final scores of $S_{\mathrm{KN}} = 0.61$ and $0.02$, respectively. These candidates fare best as KNe due to the presence of sufficient photometry to establish a luminosity, rise time, and decay rate consistent with KNe. Early rise times and steep decay rates in particular distinguish these candidates' KN scores while lowering KN-in-SN and super-KN scores. 

Two candidates (AT\,2025admd and AT\,2025ails) fare best as KNe-in-SNe. 2D scores range from $S_{\mathrm{2D}} = 0.45$ to $0.93$, and distance scores from $S_{\mathrm{dist}} = 0.14$ to $0.23$. AT\,2025admd and AT\,2025ails receive final scores of $S_{\mathrm{KN\text{-}in\text{-}SN}} = 0.13$ and $0.11$, respectively. Lightcurves are somewhat consistent with being flat or hint at slow fading or brightening, consistent with KNe-in-SNe in the time window explored and inconsistent with KNe. Whether any of these candidates brightened beyond 10 days is unknown, such that rise times are estimated as $<$$10$~days, disfavoring these candidates as super-KNe. 

Long-term photometry and spectroscopic follow-up of these candidates of interest could have clarified their nature. Further monitoring of fading candidates with notable $S_{\mathrm{KN}}$ could have better constrained decay rates to determine whether they were genuinely consistent with KNe and inconsistent with other transient types. Monitoring past 10 days could have identified whether the other candidates were brightening or displaying a plateau in their lightcurves. If this monitoring established a rise time $\geqslant$$10$~days for any of these candidates, such candidates would no longer be disfavored as super-KNe, and some might have shown $S_{\mathrm{super\text{-}KN}}$~larger than $S_{\mathrm{KN}}$ or $S_{\mathrm{KN\text{-}in\text{-}SN}}$. Moreover, three of four candidates (all but AT\,2025adin) have host galaxies with uncertain photometric redshifts. Obtaining a spectrum of the candidate or its host might have yielded a distance and potentially a classification. Unfortunately, at time of writing (March 2026), one candidate (AT\,2025admd) lies within a few degrees of the Sun and three (AT\,2025adin, AT\,2025ailp, AT\,2025ails) are at declinations $\sim$$-35^\circ$ to $-55^\circ$ and not feasibly observable with telescopes at our disposal in the Northern hemisphere. 

The behavior of these candidates and their scores with time may motivate future search and follow-up strategies for SSM GW events. If a monitored candidate retains a high $S_{\mathrm{KN\text{-}in\text{-}SN}}$ beyond $\sim$$10$~days, spectroscopic follow-up could be used to obtain a distance to the candidate or its putative host galaxy. If a candidate spectrum is obtained, this should confirm whether it is a SESN, and whether the candidate lies in the localization volume. If a SESN at a distance consistent with the GW luminosity distance, one could search for signatures of a KN embedded in the SN. The clearest signature would be the predicted NIR excess at late times, after peak, which could manifest with colors $r-J$, $r-H$, or $r-K \approx 1-2$ \citep{barnes_2022}. To search for super-KNe, if the candidate lies within the GW localization volume and monitoring yields a high score $S_{\mathrm{super\text{-}KN}}$ based on luminosities and a rise time longer than 10 days, spectra acquired near peak could be examined for broad absorption features emitted in the NIR ($\gtrsim$$10,000$~\AA; \citealt{siegel22_superKN}). These strategies are complementary: both require spectroscopic follow-up of promising candidates beyond 10 days. This follow-up would of course require late-time triggers, as are currently infrequently employed in searches for KNe only.


\section{Conclusions}\label{sec:conclusions}

We introduce a framework for ranking candidate EM counterparts to GW events involving one or more SSM objects. Taking a census of current literature, we identify KNe, KNe-in-SNe, super-KNe, and BBH merger-induced AGN flaring as the EM signatures which might accompany SSM GW events. Our framework vets candidates in search of any among this zoo of transients. We apply this new framework to S251112cm: the first statistically significant event reported by the LVK to contain at least one object with subsolar mass. We deploy a suite of telescopes to search for and follow up on candidate counterparts to S251112cm. Using our own optical photometry and the publicly available observations of the broader community, we vet and assign scores to 456 candidates with our new metrics in \texttt{TROVE}. 

Our vetting framework is designed to inform optimal follow-up and help distinguish between KNe, SESNe which might host KNe, and super-KNe, based on candidates' estimated luminosities, rise times, and decay rates. Distinguishing between classes of transient is challenging when photometry is limited or not constraining. In these cases, distance estimates are critical, but many distances come from highly uncertain photometric redshifts of candidates' potential host galaxies. Fortunately, future DESI data releases will be a massive boon to determining reliable spectroscopic redshifts. Candidates which nonetheless retain high scores due to a lack of information are high-priority targets for follow-up. These high-scoring candidates were the target of our follow up efforts, and we reported our own observations to the community in near-real time. Given the impact of limited information (\eg, uncertain distances or a dearth of constraining photometry) in setting these high scores, in future work, we will explore the possibility of quantifying the uncertainty or confidence associated with a score. At the same time, we will aim to quantify the relative importance of different subscores. 

An important result of our census of the literature is that KNe may not be the only, nor even the most likely, EM counterparts to GW events involving SSM compact objects. Searches for counterparts to these events should therefore employ strategies which consider all potential EM counterparts. Many of these counterparts evolve on longer timescales than KNe, motivating longer-term monitoring than is currently employed in search of KNe only. Regular operations of Rubin's LSST will support this sustained monitoring. However, more modeling of the transients of interest is also required, to refine photometric constraints and better define their observability and definitive signatures. KNe-in-SNe, super-KNe, and BBH-induced AGN flares are all hitherto unobserved transients. More broadly, further investigation into formation channels for ssNSs/ssBHs could identify other potential EM signatures.

Finally, we highlight that incorporating photometry from Rubin-LSST increases the number of candidate counterparts to S251112cm from 389 to 456. This dramatic increase underscores the need for efficient tools to vet candidate counterparts. With the LVK envisioning an observing run (IR1) in late 2026 and observing runs in the coming years, the era of GW follow-up supported by Rubin is imminent. To enable effective prioritization of photometric and spectroscopic follow-up in this era, the vetting framework introduced in this work will be included in the forthcoming public release of \texttt{TROVE}.

 
\begin{acknowledgments}
We thank the observing teams and technical staff of the Mt. Lemmon telescope operated under the Catalina Sky Survey, the T80-South telescope at Cerro Tololo Inter-American Observatory, Chilean and Australian telescopes of DLT40, the MMT and its Binospec spectrograph, the Southern Astrophysical Research Telescope and its Goodman spectrograph, and the Bok telescope and its Boller \& Chivens spectrograph. The Catalina Sky Survey is funded under NASA grant \#80NSSC24K1187. Time-domain research by the University of Arizona team and D.J.S. is supported by National Science Foundation (NSF) grants 2308181, 2407566, and 2432036 and the Heising-Simons Foundation under grant \#2020-1864. 

We thank the Rubin observing team for conducting a search in the localization region S251112cm in November 2025, \textit{before} Rubin entered regular operations. We are grateful to the Transients and Variable Stars - Multi-Messenger Astronomy (TVS-MMA) working group for the role they played in planning these ToO observations, which greatly benefited this work. We also thank Ish Gupta and Chris Fryer for fruitful discussions on SSM GW events and potential progenitors and EM signatures. We thank Xander Hall for helpful discussions on candidate counterparts to S251112cm. Finally, we thank the anonymous referee for their review, which has strengthened this study.

The TROVE team gratefully acknowledges support by the National Science Foundation under grant Nos. AST-2432037 and AST-2432036. N.V. gratefully acknowledges funding from the Natural Sciences and Engineering Research Council of Canada (NSERC) Postdoctoral Fellowship No. 599555. N.F. acknowledges support from the National Science Foundation Graduate Research Fellowship Program under Grant No. DGE-2137419. C.D.K. gratefully acknowledges support from the NSF through AST-2432037, the HST Guest Observer Program through HST-SNAP-17070 and HST-GO-17706, and from JWST Archival Research through JWST-AR-6241 and JWST-AR-5441. W.F. gratefully acknowledges support by the David and Lucile Packard Foundation, the Research Corporation for Science Advancement through Cottrell Scholar Award \#28284, and the National Science Foundation under grant Nos. AST-2206494, AST-2308182, and CAREER grant No. AST-2047919. J.C.R. was supported by NASA through the NASA Hubble Fellowship grant \#HST-HF2-51587.001-A awarded by the Space Telescope Science Institute, which is operated by the Association of Universities for Research in Astronomy, Inc., for NASA, under contract NAS5-26555. M.S. acknowledges funding from the Australian Research Council (ARC) Centre of Excellence CE230100016. H.B. acknowledges the financial support from ANID NATIONAL SCHOLARSHIPS/DOCTORATE 21241862. O.R. acknowledges support from the Rubin-Chile Fund under grant DIA2650, from the ANID Millennium Institute of Astrophysics (MAS) under grant ICN12\_009, and from ANID/FONDECYT grant 1251692. V.P. acknowledges support from NASA grant 80NSSC24K0771 and NSF grant PHY-2145421. A.G. acknowledges support from the Steward Observatory Fellowship in Theoretical and Computational Astrophysics, the IAU-Gruber Fellowship, and the CHE Fellowship. M.S.S. acknowledges support from the University of Zurich (UZH) and the Swiss National Science Foundation (SNSF) under grant number 10002981. J.Q.V. is supported by the European Union (ERC, Starstruck, 101095973) and the IAU-Gruber foundation. K.S. acknowledges funding from NSERC.

\end{acknowledgments}


\facilities{Bok (B\&C), CTIO:PROMPT (DLT40), Meckering:PROMPT (DLT40), MMT (Binospec), SO:1.5m (CSS); SOAR (Goodman), S-PLUS (T80-South)}

\software{Astropy \citep{astropy1, astropy2, astropy3},
\texttt{astropy-healpix} \citep{astropy3},
\texttt{astroquery} \citep{ginsburg_astroquery_2019},
Astro-SCRAPPY \citep{mccully_astropy_2018},
Beautiful Soup \citep{richardson_beautiful_2023},
Binospec IDL Pipeline \citep{Kansky2019},
CASA \citep{mcmullin_casa_2007, casa_team_casa_2022},
\texttt{crispy-bootstrap4} \citep{smith_crispy-bootstrap4_2022},
\texttt{dateutil} \citep{niemeyer_dateutil_2021},
Django \citep{django_software_foundation_django_2023},
\texttt{django-bootstrap4} \citep{verheul_django-bootstrap4_2021},
\texttt{django-crispy-forms} \citep{araujo_django-crispy-forms_2023},
Django Extensions \citep{trier_django_2023},
Django Filter \citep{gibson_django_2021},
\texttt{django-gravatar} \citep{waddington_django-gravatar_2020},
\texttt{django-guardian} \citep{balcerzak_django-guardian_2021},
Django REST Framework \citep{christie_django_2022},
\texttt{django-webpack-loader} \citep{lone_django-webpack-loader_2022},
\texttt{dustmaps} \citet{2018JOSS....3..695M},
\texttt{fastavro} \citep{tebeka_fastavro_2022},
Flask \citep{pallets_projects_flask_2022},
Flask-SQLAlchemy \citep{pallets_projects_flask-sqlalchemy_2021},
\texttt{fundamentals} \citep{young_fundamentals_2023},
\texttt{gracedb-sdk} \citep{singer_gracedb-sdk_2022},
HEALPix Alchemy \citep{singer_healpix_2022},
\texttt{healpy} \citep{zonca_healpy_2019},
Hop Client \citep{godwin_hop_2022},
Hopskotch \citep{scimma_project_hopskotch_2023},
\texttt{IRAF} \citep{Tody1986, Tody1993},
Light Curve Fitting \citep{hosseinzadeh_light_2022},
\texttt{ligo.skymap} \citep{singer_rapid_2016},
\texttt{lmfit} \citep{newville_lmfit_2023},
Apache Kafka \citep{apache_software_foundation_apache_2023},
Matplotlib \citep{hunter_matplotlib:_2007},
MOCPy \citep{baumann_cds-astro_2023},
NumPy \citep{harris_array_2020},
\texttt{paramiko} \citep{forcier_paramiko_2023},
\texttt{photutils} \citep{bradley_astropy_2022},
\texttt{pillow} \citep{murray_python-pillow_2023},
\texttt{plotly.py} \citep{plotly_plotly_2023},
PostgreSQL \citep{postgresql_global_development_group_postgresql_2022},
\text{psycopg} \citep{di_gregorio_psycopg_2023},
Python-Markdown \citep{stienstra_python-markdown_2023},
Q3C \citep{koposov_q3c_2006},
\texttt{redback} \citep{sarin_redback_2024},
\texttt{requests} \citep{reitz_requests_2023},
SAGUARO Pipeline \citep{2023zndo...8436113P},
SAGUARO TOM \citep{hosseinzadeh_saguaro_2023},
SASSy Q3C Models \citep{daly_sassy_2023},
SciPy \citep{virtanen_scipy_2020},
\texttt{setuptools} \citep{python_packaging_authority_setuptools_2023},
\texttt{sip\_tpv} \citep{shupe_more_2012},
Source Extractor \citep{bertin_sextractor:_1996,bertin_sextractor_2010},
\texttt{SNID-SAGE} \citep{snid_sage_2025}, 
\texttt{specutils} \citep{earl_astropy_2023},
SQLAlchemy \citep{bayer_sqlalchemy_2012},
SWarp \citep{swarp},
TOM Toolkit \citep{collom_tom_2020,lindstrom_tom_2022},
\texttt{tom-alertstreams} \citep{tom_toolkit_project_tom-alertstreams_2023},
\texttt{tom\_nonlocalizedevents} \citep{tom_toolkit_project_tom_nonlocalizedevents_2023},
\texttt{urllib3} \citep{petrov_urllib3_2023},
\texttt{voevent-parse} \citep{staley_voevent-parse_2014},
Watchdog \citep{mangalapilly_watchdog_2023}
}


\bibliography{main, GCNs_S251112cm, TNS}{}
\bibliographystyle{aasjournalv7}


\appendix


\section{Vetting for BBH Merger-Induced AGN Flaring}\label{app:BBHAGN}

When searching for BBH-induced AGN flaring, predetection at the location of a candidate is not disqualifying. The nature of a predetected source is already captured by the combination of point source, minor planet, and AGN association. We therefore simply allow predetection at any time pre-merger when vetting for AGN flares.

Unfortunately, it is challenging to impose robust constraints on the luminosities and timescales of BBH-induced AGN flares. Differences in behavior arise in part from differing emission mechanisms, as rapid accretion onto the kicked remnant BH may power relativistic jets and/or disk winds. These jets or winds may subsequently interact with the surrounding disk and produce observable EM radiation. In both cases, flare luminosity is expected to scale approximately as $L \propto M_{\mathrm{BBH}}^{2}\,v_{\mathrm{k}}^{-3}\,\rho$, where $M_{\mathrm{BBH}} = m_1 + m_2$ is the total binary mass, $v_{\mathrm{k}}$ the kick velocity, and $\rho$ the ambient gas density in the AGN disk. Different emission mechanisms then predict soft X-ray emission on timescales of weeks \citep{Kimura2021}, while several others \citep{mckernan19_BBHAGN,  RodrguezRamrez2023, Tagawa2024, rodriguez25_AGNflare} predict UV/optical flares lasting from weeks to several hundred days. Isolating just one mechanism (disk winds), \citet{rodriguez25_AGNflare} explore parameter space by varying kick velocity ($v_{\mathrm{k}} = 200$--$400~\mathrm{km~s^{-1}}$) and SMBH mass ($M_{\mathrm{SMBH}} = 10^{6}$--$10^{7}~M_{\odot}$), while fixing $M_{\mathrm{BBH}} = 20~M_{\odot}$ and the kick angle at $\theta_k = 10^{\circ}$ from the disk midplane. They find peak times in the range of $\sim$$10-80$~days and peak luminosities of $\sim 5 \times 10^{42}$ to $5 \times 10^{43}~\mathrm{erg~s^{-1}}$. It has also been shown that flares are broadly classified into two categories based on their delay times relative to the merger \citep{darc25_BBHflares}. Short-delay flares are expected to occur within $\lesssim 50$ days of the merger and are typically associated with relatively short durations. In contrast, long-delay flares can peak between $\sim$$50$ and $\sim$$400$ days, and last longer. 

Candidates are thus scored based only on their 2D position in the GW localization region, the absence of point sources or minor planets, 2D association with an AGN, and the distance to the AGN. To search for AGN flares among the candidates, one may select candidates with non-zero $S_{\mathrm{AGN}\text{-}\mathrm{flare}}$ and examine photometry for clear evidence of flaring above regular AGN variability. This type of modeling has been conducted for select BBH GW events \citep{graham20_GW190521flare, cabrera24_S230922g_AGN, darc25_BBHflares, bommireddy26}. Here, we search for AGN flaring only among candidate transients already identified with optical/IR/X-ray observations. We will explore a more holistic approach in the future that accounts for all known AGN within a region, regardless of their multi-wavelength behavior, similar to other works (\eg, \citealt{graham23_BBH-EM, darc25_BBHflares, zhu26_AGNflares_ZTF, mcpike26_McFACTSIV, bommireddy26}).


\section{Detailed Scores and Inferred Photometric Quantities}\label{app:scores-all}

We present here the scores and inferred photometric quantities which are obtained from scoring all 456 objects reported to the TNS or in GCNs within 10 days of the GW signal. Table~\ref{tab:scores_allcands}~shows, for all candidates: the 2D localization subscore $S_{\mathrm{2D}}$, distance subscore $S_{\mathrm{dist}}$, point source and minor planet association score factors $S_{\mathrm{PS}}$ and $S_{\mathrm{MPC}}$, AGN association subscore $S_{\mathrm{AGN}}$, and photometry subscores $S_{\mathrm{phot}}$. Finally, we list the overall scores $S_T$ for transient class $T$, among KNe, KNe-in-SNe, and super-KNe. We omit scores for BBH-induced AGN flaring as only two candidates are associated with AGN. Distance, point source, minor planet, and AGN scores are the same when vetting for KNe, KNe-in-SNe, and super-KNe, for a given candidate. 

Table~\ref{tab:photom_allcands} shows the peak luminosity $L_{\mathrm{peak}}$, rise time to peak $t_{\mathrm{peak}}$, and decay rate $\Delta m$ (where a positive value indicates fading) inferred for all candidates. We also indicate whether a candidate is considered detected prior to the GW signal. Where a quantity is missing, available photometry is insufficient to constrain it. By comparing these inferred quantities to observation and theory-driven photometry metrics in Table~\ref{tab:photom-metrics} (Section~\ref{ssc:scoring-photom}), we compute score factors which ultimately yield the photometry subscore in Table~\ref{tab:scores_allcands}. Whether a candidate is considered predetected, its rise time $t_{\mathrm{peak}}$, and its decay rate $\Delta m$ are dependent on the window over which we fit photometry: up to 25 days for KNe, and up to 100 days for KNe-in-SNe and super-KNe.

\startlongtable



\section{Candidates of Interest}\label{app:interest}

\subsection{Candidates With Notable KN or KN-in-SN Scores}\label{sssc:interesting-KN-KN-in-SN-super-KN}

\begin{figure*}[!ht]
    \centering
    \includegraphics[width=\linewidth]{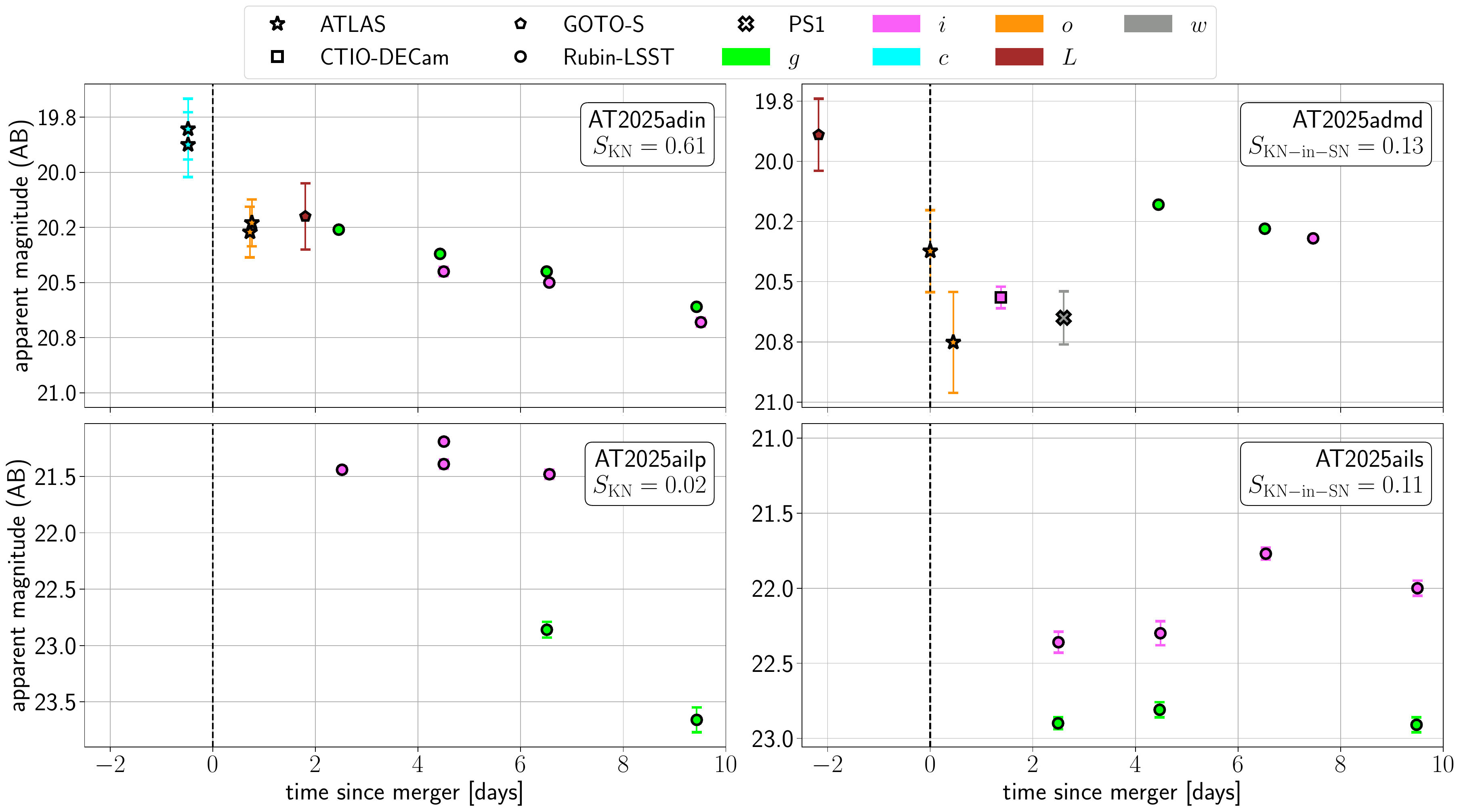}
    \includegraphics[width=\linewidth]{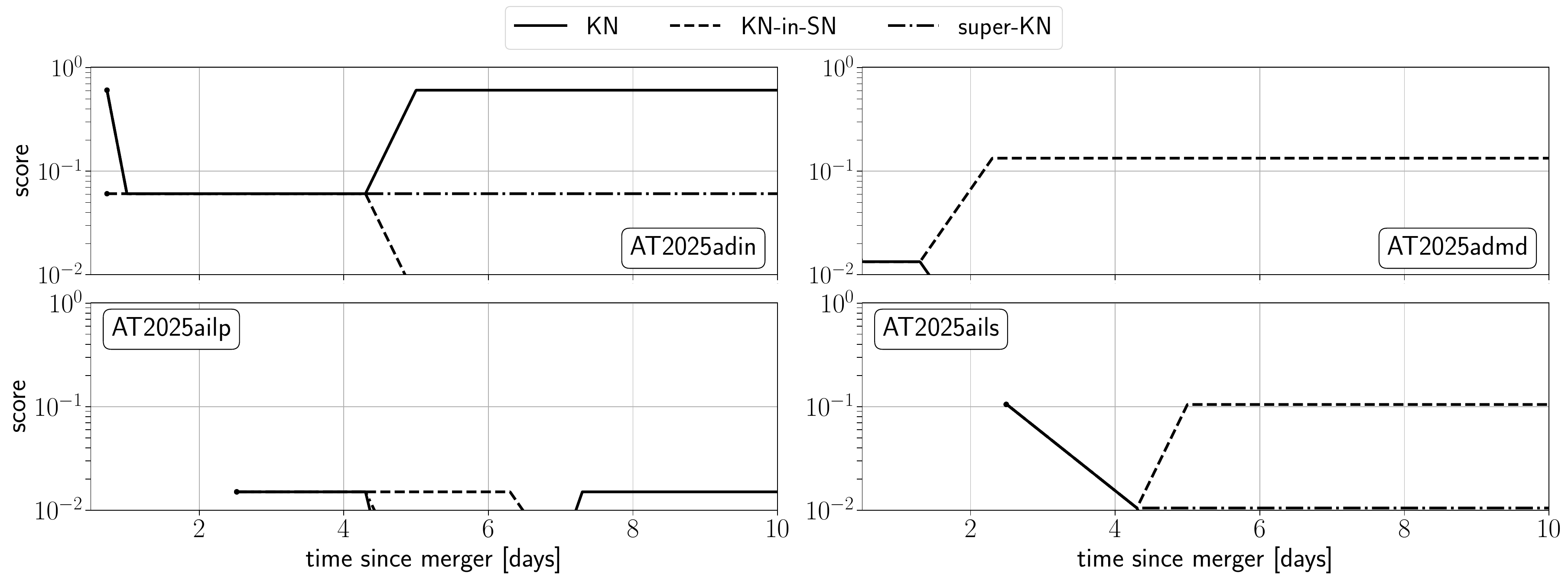}  \caption{\textbf{Optical lightcurves and scores with time of four candidates of interest.} \textit{Top:} We annotate each candidate with its KN or KN-in-SN score. Lightcurves are constructed from public photometry from DECam \citep{gcn42691}, GOTO-South \citep{gcn42658}, Rubin-LSST \citep{gcn43257}, PS1, and ATLAS forced photometry \citep{tonry18_ATLAS, smith20_ATLAS, shingles21_ATLAS}. Different markers denote different instruments, while different colors indicate photometric band. For some Rubin photometry, uncertainties are on the order of point size. \textit{Bottom}: Corresponding KN (solid), KN-in-SN (dashed), and super-KN (dash-dotted) scores with time. Each line, beginning at the point of first $\geqslant$$5\sigma$ optical detection, describes a candidate's score with time as new photometry is obtained. Scores with time are plotted from 0.5 to 10 days.}
    \label{fig:interesting-lightcurves}
\end{figure*}

We highlight four candidates of interest with notable KN or KN-in-SN scores and sufficient photometry to distinguish between scores for different transient types. No candidates fare best as super-KNe, which we ascribe to counterpart search strategies or simply the absence of super-KNe in nature. We show the lightcurves and scores with time of these candidates in Figure~\ref{fig:interesting-lightcurves}. We discuss each candidate in turn below.

\textit{\textbf{AT\,2025adin}}: AT\,2025adin receives $S_{\mathrm{2D}} = 0.70$ and $S_{\mathrm{dist}} = 0.87$ from its association with a galaxy with $z$-independent distance $77.3 \pm 17.0~\mathrm{Mpc}$ (GWGC; \citealt{White+11_GWGC}).\footnote{\citet{gcn43257} note that AT\,2025adin is within 50 kpc of a NED-LVS galaxy; we find the galaxy WISEA J020404.81-513833.9 offset by 24'' ($\sim$$10.5$ kpc at $\sim$$93$ Mpc) with $z = 0.01943 \pm 0.00015$. The GWGC galaxy we identify is offset by 24'' and likely corresponds to the same object.} Its rapid decline (also noted in \citealt{gcn43257}) disfavors it as a KN-in-SN and its correspondingly short rise time disfavors it as a super-KN. While marginally detected in ATLAS $c$-band just before merger, we require three $\geqslant$$5\sigma$ detections in a $\pm 5$-day window to declare a source predetected, and only two are made. The association with a distant galaxy, albeit with an uncertain photometric $z$, almost certainly means that AT\,2025adin is not associated with S251112cm. We present it nonetheless due to its notable decline in multiple bands and as an example of our framework's ability to distinguish between KNe, KNe-in-SNe, and super-KNe. AT\,2025adin demonstrates the importance of monitoring, as its KN and KN-in-SN scores swap values at 4.5 days once enough Rubin-LSST $g$- and $i$-band photometry has been collected to establish a decay rate too rapid for a KN-in-SN. It receives a final $S_{\mathrm{KN}} = 0.61$.

\textit{\textbf{AT\,2025admd}}: AT\,2025admd receives $S_{\mathrm{2D}} = 0.98$ and $S_{\mathrm{dist}} = 0.15$ from its association with a galaxy with photometric $z = 0.11^{+0.03}_{-0.04}$ (LS DR10). It is detected in GOTO-$L$ at 2 days pre-merger, but this one detection is not sufficient to establish the candidate as pre-detected. It is immediately disfavored as a KN or super-KN when photometry at $\sim$$1$~day post-merger establishes its luminosity as $> 10^{43}~\mathrm{erg~s^{-1}}$. It continues to brighten, reaching $L_{\mathrm{peak}} \approx  2.1 \times  10^{43}~\mathrm{erg~s^{-1}}$, based on Rubin-LSST $g$-band photometry at 4.5 days. Like AT\,2025adkm, the slower photometric evolution evident as of 4.5 days post-merger favors it as a KN-in-SN and disfavors it as a KN. Finally, the rise time estimate never exceeds 10 days, disfavoring the candidate as a super-KN. Only the KN-in-SN score survives, attaining $S_{\mathrm{KN}\text{-}\mathrm{in}\text{-}\mathrm{SN}} = 0.13$ and retaining this score. However, as with AT\,2025adkm, the largely flat lightcurve may be consistent with a slow-evolving SN II-P unrelated to S251112cm.

\textit{\textbf{AT\,2025ailp}}: AT\,2025ailp receives $S_{\mathrm{2D}} = 0.06$ and $S_{\mathrm{dist}} = 0.23$ from its association with a galaxy with photometric $z = 0.14^{+0.05}_{-0.08}$ (LS DR10). While initially appearing flat in the $i$-band from 2.5 to 4.5 days, reducing its KN score, it then fades by $\sim$$0.3~\mathrm{mag~day^{-1}}$ in the $g$-band from 6 to 10 days. This transition from appearing flat to fading results in the KN-in-SN score dropping while the KN score rises to and retains $S_{\mathrm{KN}} = 0.02$. This is the lowest score among our highlighted candidates of interest, but once again demonstrates the impact of collecting photometry.

\textit{\textbf{AT\,2025ails}}: AT\,2025ails receives $S_{\mathrm{2D}} = 0.45$ and $S_{\mathrm{dist}} = 0.23$ from its association with a galaxy with photometric $z = 0.18^{+0.13}_{-0.14}$ (LS DR10). $g$- and $i$-band photometry hint at brightening and a rise time inconsistent with a KN. The lack of photometry beyond 10 days is such that the rise time is still estimated as $< 10~\mathrm{days}$, also disfavoring the candidate as a super-KN. Only the KN-in-SN score survives, attaining $S_{\mathrm{KN}\text{-}\mathrm{in}\text{-}\mathrm{SN}} = 0.11$ and retaining this score. 

We also note SN\,2025adtq: a SN\,IIb highlighted in \citet{hall26_S251112cm}. It receives $S_{\mathrm{2D}} = 0.26$ based on its position in the localization region and $S_{\mathrm{dist}} = 0.32$ based on its measured redshift of $z = 0.033$. Based on publicly available photometry, it rises at a later time than expected for KNe. It ultimately receives $S_{\mathrm{super}\text{-}\mathrm{KN}} = 0.08$.

\subsection{Candidates Associated With AGN}\label{ssc:interest-2025adnc-and-X3}

Six candidates are associated with AGN: AT\,2025adid, AT\,2025adnc, AT\,2025adpr, AT\,2025adps, AT\,2025adrz, and S251112cm\_X3. This association sets $S_{\mathrm{KN}} = S_{\mathrm{KN\text{-}in\text{-}SN}} = S_{\mathrm{super\text{-}KN}} = 0$ for \red{all six} candidates, while $S_{\mathrm{AGN\text{-}flare}} > 0$. However, all but S251112cm\_X3 are associated with AGN confidently outside of the localization volume, such that S251112cm\_X3 is the only candidsate with $S_{\mathrm{AGN\text{-}flare}} \geqslant 0.01$. 

S251112cm\_X3 is a known X-ray source, and shows no X-ray outburst near the time of S251112cm, but is nonetheless reported due to its position in the localization region (\citealt{gcn42681}). It receives $S_{\mathrm{2D}} = 0.62$. The AGN associated with the candidate, MCG $-$01.24.012, has a redshift $z = 0.02$ (Milliquas; \citealt{flesch23_milliquasv8}; fiducial uncertainty $0.01z = \pm 0.0002$), yielding a distance score $S_{\mathrm{dist,AGN\text{-}flare}} = 0.21$. S251112cm\_X3 thus finally receives a score $S_{\mathrm{AGN}\text{-}\mathrm{flare}} = 0.13$. This is the highest (the only truly non-zero) score $S_{\mathrm{AGN}\text{-}\mathrm{flare}}$ among all candidates. But, ATLAS forced photometry largely finds no optical counterpart down to $o$ and $c \sim 21$ at all times, except for marginal ($\sim$$5\sigma$) detections of $o \approx 19.6 \pm 0.1$ at 19, 31, and 32 days and $o \approx 20.3 \pm 0.2$ at 67 days. These may be the result of poor subtraction given the proximity of the candidate to the nucleus of the galaxy, and no other UV/optical photometry exists. Deeper and higher-resolution UV/optical photometry could have more clearly identified any transient behavior, and whether any flaring was consistent with regular AGN variability or a BBH merger in the AGN disk. As it stands, the ATLAS photometry is inconclusive.

\end{document}